\documentclass{ws-ijqi}
\usepackage[super,sort,compress]{cite}
\usepackage{etoolbox}
\usepackage{color,bm,bbold}
\usepackage[x11names]{xcolor}
\usepackage[colorlinks=true,citecolor=red,linkcolor=blue,urlcolor=SpringGreen4]{hyperref}

\newcommand{\ket}[1]{|{#1}\rangle}
\newcommand{\bra}[1]{\langle{#1}|}
\newcommand{\I}{\mathrm{i}}
\newcommand{\D}{\mathrm{d}}
\newcommand{\ML}[1]{\widehat{#1}_{\textsc{ml}}}
\newcommand{\LS}[1]{\widehat{#1}_{\textsc{ls}}}
\newcommand{\inner}[2]{\langle{#1}|{#2}\rangle}
\newcommand{\opinner}[3]{\langle{#1}|{#2}|{#3}\rangle}

\newcommand{\rvec}[1]{\bm{#1}}
\newcommand{\dyadic}[1]{\bm{#1}}
\newcommand{\tr}[1]{\mathrm{tr}\{#1\}}
\newcommand{\Tr}[1]{\mathrm{Tr}\{#1\}}
\newcommand{\ptr}[2]{\mathrm{tr}_{#1}\!\left\{#2\right\}}
\newcommand{\sket}[1]{|{#1}\rangle\!\rangle}
\newcommand{\sbra}[1]{\langle\!\langle{#1}|}
\newcommand{\sinner}[2]{\langle\!\langle{#1}|{#2}\rangle\!\rangle}
\newcommand{\opsinner}[3]{\langle\!\langle{#1}|{#2}|{#3}\rangle\!\rangle}
\newcommand{\ssubset}{\subset\joinrel\subset}
\newcommand{\openone}{\mathbb{1}}

\makeatletter
\def\@mkboth#1#2{}
\newlength\appendixwidth
\preto\appendix{\addtocontents{toc}{\protect\patchl@section}}
\newcommand{\patchl@section}{%
	\settowidth{\appendixwidth}{\textbf{Appendix }}%
	\addtolength{\appendixwidth}{1.5em}%
	\patchcmd{\l@section}{1.5em}{\appendixwidth}{}{\ddt}%
}
\makeatother

\newtheorem{prop}[lemma]{Property}

\newcounter{algoctno}
\newcommand{\algoct}{%
	\stepcounter{algoctno}%
	\thealgoctno}

\newcounter{algolabno}
\newcommand{\algolab}[1]{\refstepcounter{algolabno}\label{#1}}

\begin{document}

\def\bib{B\kern-.05em{I}\kern-.025em{B}\kern-.08em}
\def\btex{B\kern-.05em{I}\kern-.025em{B}\kern-.08em\TeX}

%%%%%%%%%%%%%%%%%%%%% Publisher's Area please ignore %%%%%%%%%%%%%%
\catchline{}{}{}{}{}
%%%%%%%%%%%%%%%%%%%%%%%%%%%%%%%%%%%%%%%%%%%%%%%%%%%%%%%%%%%%%%%%%%%

\title{MODERN COMPRESSIVE TOMOGRAPHY FOR QUANTUM INFORMATION SCIENCE}

\author{YONG SIAH TEO}

\address{Department of Physics and Astronomy, Seoul National University, 08826 Seoul, South Korea\\
ys\_teo@snu.ac.kr}

\author{LUIS L. S{\'A}NCHEZ-SOTO}

\address{Departamento de \'Optica, Facultad de F\'{\i}sica,
Universidad Complutense, 28040 Madrid, Spain \\
Max-Planck-Institut f\"ur  die Physik des Lichts,
Staudtstra\ss e 2, 91058 Erlangen, Germany \\ 
lsanchez@fis.ucm.es}

\maketitle

%\begin{history}
%\received{Day Month Year}
%\revised{Day Month Year}
%%\accepted{Day Month Year}
%%\comby{(xxxxxxxxxx)}
%\end{history}

\begin{abstract}
This review serves as a concise introductory survey of modern compressive tomography developed since 2019. These are schemes meant for characterizing arbitrary low-rank quantum objects, be it an unknown state, a process or detector, using minimal measuring resources (hence compressive) without any \emph{a priori} assumptions (rank, sparsity, eigenbasis, \emph{etc}.) about the quantum object. This article contains a reasonable amount of technical details for the quantum-information community to start applying the methods discussed here. To facilitate the understanding of formulation logic and physics of compressive tomography, the theoretical concepts and important numerical results (both new and cross-referenced) shall be presented in a pedagogical manner.
\end{abstract}

\keywords{compressive tomography; quantum tomography; compressed sensing.}

\tableofcontents  % optional

\markboth{Yong Siah Teo and Luis L. S{\'a}nchez-Soto}
{Modern compressive tomography for quantum information science}

\section*{Acronyms}
\addcontentsline{toc}{section}{Acronyms}

\begin{tabular}{rl}
	ACT & adaptive compressive (state) tomography\\
	ACTQPT & ACT-based QPT (input-state-output-POVM formalism)\\
	ACQPT & adaptive compressive QPT\\
	BF & Baldwin--Flammia (POVM)\\
	BG & Baldwin--Goyeneche (bases)\\
	BKD & Baldwin--Kalev--Deutsch\\
	BM & Bayesian mean\\	
	CP & completely-positive\\
	CQDT & compressive QDT\\
	CQPT & compressive QPT\\
	CQST & compressive QST\\
	CS & compressed sensing\\
	CS-QST & CS-based QST\\
	IC & informationally complete\\
	ICC & informational completeness certification\\
	KW & Kech--Wolf\\
 	LS & least squares\\
 	ML & maximum likelihood\\
 	NISQ & noisy intermediate-scale quantum (devices)\\
 	PACT & product ACT\\
 	PACTQPT & product ACTQPT\\
	POVM & positive operator valued measure\\	
	QDT & quantum detector tomography\\
	QMT & quantum metrology\\
	QPT & quantum process tomography\\
	QST & quantum state tomography\\	
	RH & random Haar (bases)\\
	RLH & random local Haar (bases)\\
	RIP & restricted isometry property\\
	SDP & semidefinite program\\
	SVD & singular-value decomposition\\
	TP & trace-preserving
\end{tabular}

\newpage

\allowdisplaybreaks

\section*{Introduction}
\addcontentsline{toc}{section}{Introduction}

A practically feasible toolkit for characterizing quantum states~\cite{Chuang:2000fk,lnp:2004uq,Teo:2015qs}, processes (or channels)~\cite{Chuang:2000fk,OBrien:2004aa,Fiurasek:2001dn,Poyatos:1997aa,Teo2011aa} and detectors~\cite{Luis:1999qm,Fiurasek:2001mq,D'Ariano:2004aa,Chen:2019aa,Zhang:2012aa,Altorio:2016aa,Motka:2017aa} is a necessity for building devices~\cite{Ladd:2010aa,Campbell:2017aa,Lekitsch:2017aa} that carry out quantum-information and computation tasks reliably. As quantum devices grow in size, standard tomographic procedures for fully reconstructing complete operators that represent the components in such devices typically follow a polynomial complexity with respect to the relevant Hilbert-space dimension. Resource-efficient tomography schemes are therefore required.

An attempt to establish such efficient \emph{quantum-state tomography}~(QST) schemes is based on the premise that the unknown quantum state to be characterized is rank-deficient with a rank no larger than some integer $r$, where $r$ is assumed to be known. This is essentially the basis-independent version of the notion of sparsity in signal processing---a rank-$r$ quantum state is $r$-sparse in its unique eigenbasis. This connection mathematically invited the concept of compressed sensing~(CS), which was first introduced to perform compressive signal/image reconstruction~\cite{Donoho:2006cs,Candes:2006cs,Candes:2009cs}. This framework was then later borrowed and implemented in QST~\cite{Gross:2010cs,Kalev:2015aa,Baldwin:2016cs,Steffens:2017cs,Riofrio:2017cs}, where it was found that only around $O(rd)\ll O(d^2)$ measurements are needed to uniquely reconstruct any unknown state of Hilbert-space dimension $d$. The framework of CS has also been applied to \emph{quantum process tomography}~(QPT)~\cite{Baldwin:2014aa,Rodionov:2014aa,Shabani:2011aa}.

As we shall witness in Sec.~\ref{sec:CS}, such CS-based QST protocols are met with shortcomings. Firstly, the purported rank assumption, that is the knowledge of $r$, is always difficult to justify in real experimental situations, where noise generally increases $r$ unbeknownst to the experimental observer. Using the wrong $r$ value results in the erroneous choice of compressive measurements that would lead to an unreliable reconstruction. Any additional means that would help certify the value of $r$ would hence need to be considered as part of the CS scheme, thereby altering the $O(rd)$ scaling. Secondly, given the chosen compressive measurements and data, these schemes do not offer a systematic verification of whether the resulting reconstruction is indeed unique. Rather, the quantum fidelity is a favorite measure of choice for judging its quality~\cite{Kalev:2015aa,Steffens:2017cs,Riofrio:2017cs}. Obviously, the fidelity is unable to tell us whether the reconstruction is unique or not, which is the main question in compressive tomography, unless its value is unity and this is never possible in noisy experiments. Thirdly, provable CS-QST measurements with known tight scaling for $r\ll d$ are restricted to highly specific measurements, two examples include the random Pauli-observable expectation~\cite{Gross:2010cs} and element-probing measurements~\cite{Baldwin:2016cs}.

Since 2019, we reported several works~\cite{Ahn:2019aa,Ahn:2019ns,Teo:2020cs,Kim:2020aa,Gianani:2020aa,teo2021benchmarking,gillopez2021universal} that investigate the problem of compressive tomography in different perspectives that are completely independent from CS. Essentially, we developed a self-consistent \emph{informational completeness certification}~(ICC) routine that can decisively state whether a given set of measurements and data is \emph{informationally complete}~(IC); that is, whether a unique reconstruction of the unknown quantum object is possible. No assumptions (including rank, sparsity, eigenbasis, \emph{etc.}) whatsoever about the quantum object are necessary. Using such objective certification protocols, we demonstrated that even generic measurements are compressive if the unknown quantum object has low rank. For QST with random von~Neumann bases~\cite{Ahn:2019aa,Ahn:2019ns,gillopez2021universal}, for instance, our ICC protocol numerically identifies an average scaling behavior in the minimum number of IC basis measurements needed to reconstruct a rank-$r$ state that is tighter than the necessary scaling previously derived by Kech and Wolf~\cite{Kech:2016aa}. Applications to QPT~\cite{Teo:2020cs,Kim:2020aa} were also subsequently established. The versatility of ICC allows us to universally carry out compressive tomography on any low-rank quantum object, which permits us to also perform \emph{quantum detector tomography}~(QDT) compressively~\cite{Gianani:2020aa}, at which point no equivalent CS-based schemes existed to the authors' knowledge. Furthermore, we managed to establish \emph{adaptive} methods~\cite{Ahn:2019aa,Ahn:2019ns,Kim:2020aa} that require less optimal measurements than random ones, thereby increasing the compressivity in quantum tomography. Another important observation is that these modern compressive methods can generally be carried out with local measurement resources, which makes them ideal for implementation on noisy intermediate-scale quantum~(NISQ) devices.

These matters shall be reported in this review, beginning with Sec.~\ref{sec:CS}, where we briefly outline the procedure of CS-QST and forewarn the reader of the pitfalls in these procedures. Next, in Sec.~\ref{sec:modCT}, we discuss basic concepts leading to the formulation of our so-called ``modern compressive tomography'' umbrella (works dating from 2019 onwards), under which arbitrary low-rank quantum objects can be compressively characterized without any spurious or \emph{ad hoc} assumptions. Sections~\ref{sec:CQST}, \ref{sec:CQPT} and \ref{sec:CQDT}, thereafter, shall provide an excursion to all recently established compressive tomography schemes for quantum states, processes and detectors. We finally conclude matters in Sec.~\ref{sec:conc}. While most graphical results pertain to noiseless scenarios for the purpose of illustrating the key scaling behaviors of such compressive schemes, we encourage the reader to consult the cited original works for all experimental results.

\section{Compressed-sensing state tomography: a brief summary}
\label{sec:CS}

The procedure of CS was formulated in the context of sparse signal recovery~\cite{Donoho:2006cs,Candes:2006cs,Candes:2009cs}, where in some practical instances, especially in the problem of image recovery, the acquired signal is sparse and the degree of sparsity $s\ll D$ of a $D$-dimensional signal column is generally assumed to be well-known. Here, an $s$-sparse signal is represented by a column with at most $s$ nonzero values. That both the value of $s$ and the basis representing this $s$-sparse signal are known is taken for granted. By choosing a specialized set of measurements (such as random ones drawn from Gaussian or Bernoulli distributions~\cite{Candes:2006cs}, or random frequency measurements in the Fourier plane for recovering a sparse temporal signal~\cite{Candes:2006bb}) and applying an optimization routine to the resulting measured signal data, it was proven that the number of measurements required to reconstruct an $s$-sparse signal scales linearly in $s$. 

A sufficient condition for measurements to be compressive under the CS framework is the so-called \emph{restricted isometry property}~(RIP). Apart from a few classes of measurements known to possess the RIP, verifying whether a given set of measurements has RIP was shown to be NP-hard~\cite{Tillmann:2006aa}. It has also been found~\cite{Candes06stablesignal} that CS still works for noisy measurements in the mathematical sense that the recovered signal is close to the true one \emph{provided that the noise level is sufficiently small}. On the other hand, certain RIPless measurements can still be compressive, but are otherwise less resilient to noise~\cite{KUENG2014110}.

The CS procedure was later ported over to QST~\cite{Gross:2010cs,Schwemmer:2014aa,Steffens:2017cs,Riofrio:2017cs} as a means to reduce the measurement resources needed to reconstruct a rank-$r$ state $\rho_r$ specified by $(2d-r)r-1$ independent parameters~(see \ref{app:rank-r}). We note that the state rank is the basis-independent version of sparsity, in the sense that a $d$-dimensional $\rho_r$ is indeed representable as a $d$-dimensional $r$-sparse column containing no more than $r$ nonzero positive eigenvalues if so desired. For the purpose of CS-QST, the authors in the aforementioned references considered a set $\{O_1,O_2,\ldots,O_M\}$ of $M\ll 4^n$ \emph{randomized Pauli observable} measurements, where each $O_m$ is an $n$-qubit tensor product of the single-qubit operators $\{\openone,\sigma_x,\sigma_y,\sigma_z\}$ with $\sigma_j$ being the standard Pauli operator in the $j$-axis. These random Pauli observables are measured with respect to an ``unknown'' $n$-qubit state $\rho_r$ ($d=2^n$) of \emph{known} $r\geq\mathrm{rank}\{\rho_r\}$. Suppose that after the \emph{noiseless} measurement results $\{o_{m}=\tr{\rho_r O_m}\}^M_{m=1}$ are collected, 
one performs the following convex-optimization algorithm:
\begin{equation}
\mathrm{minimize}\,\,\,\|\rho\|_1\,\,\,\mathrm{subject\,\, to:}\,\,\,\tr{\rho'}=1\,\,\,\mathrm{and}\,\,\,\tr{\rho'O_m}=o_m\,\,\mathrm{for}\,\,1\leq m\leq M\,,
\end{equation}
where $\|\cdot\|_1$ is the Schatten 1-norm or trace-class norm, then it can be shown that only $M=M_\textsc{ic}=O(rd\log^2 d)$ IC Pauli-observable measurements are enough to recover $\rho_r$ with very high probability. In principle, $M_\textsc{ic}\ll d^2$ if $r\ll d$. What if the data collected are noisy---$o_{m}\neq\tr{\rho_r O_m}$? Then \textbf{Observation~1} in~Ref~\cite{Gross:2010cs} guarantees a recovered state estimator that is close to $\rho_r$ provided that the noise level in the measured data is \emph{sufficiently small}.

There exists a separate project~\cite{Baldwin:2016cs} that constructed deterministic compressive measurements \emph{with the assumption that} $r\geq\mathrm{rank}\{\rho_r\}$ \emph{of known} $r$. These constructions pay no attention to the RIP, which demonstrates that RIP is not necessary for compression. More specifically, these measurements are element-probing~(EP); each measurement outcome probes a subset of elements in the matrix representation of $\rho_r$. If a generalized measurement or positive operator-valued measure~(POVM) is an option, upon denoting the standard basis as $\{\ket{l}\}$, the following non-projective Baldwin--Flammia~(BF) POVM,
\begin{align}
\Pi_{0\leq l\leq r-1}\propto&\,\ket{l}\bra{l}\,,\nonumber\\
\Pi_{r\leq l\leq (2d-r+1)r/2-1}\propto&\,(\openone+\ket{l'}\bra{m'}+\ket{m'}\bra{l'})\,\,\,\mathrm{for}\,\,\begin{cases}l'+1\leq m'\leq d-1\\0\leq l'\leq r-1\end{cases},\nonumber\\
\Pi_{(2d-r+1)r/2\leq l\leq (2d-r)r-1}\propto&\,(\openone-\ket{l'}\I\bra{m'}+\ket{m'}\I\bra{l'})\,\,\,\mathrm{for}\,\,\begin{cases}l'+1\leq m'\leq d-1\\
0\leq l'\leq r-1\end{cases},\nonumber\\
\Pi_{(2d-r)r}=&\,\openone-\sum^{(2d-r)r-1}_{l=0}\Pi_l\,,
\end{align}
contains $M_\textsc{ic}=(2d-r)r+1$ outcomes that can uniquely characterize $\rho_r$ of known $r$, which generalizes previous studies on pure-state tomography~\cite{Flammia:2005fk}. If one is restricted to von~Neumann bases, then the alternative Baldwin--Goyeneche~(BG) construction containing $K_\textsc{ic}=4r+1$ bases guarantees a unique reconstruction. This construction is slightly more complicated, hence we direct the interested reader to \textbf{Algorithm~1} of the relevant reference article~\cite{Baldwin:2016cs} for more details. In the same article, it was also shown that using one of the two constructions and the maximum-likelihood~(ML) method~\cite{Aldrich:1997ml,Banaszek:1999ml,Rehacek:2007ml,Teo:2011me,Teo:2012cs,Teo:2012ve,Shang:2017sf} can recover $\rho_r$. Again, \emph{with sufficiently small data noise}, one could obtain a state estimator that is near $\rho_r$. 

All discussions in this section, therefore, share a common consensus: \emph{so long as $r$ is known with high credibility and the data noise level is sufficiently small, CS-related schemes are reliable.} Thanks to this putative consensus, CS-based protocols has since been the go-to \emph{status quo} solutions for resource-efficient tomography. We shall now forewarn the reader of the pitfalls in relying on this consensus. First, notice that whichever CS-based procedure is employed, the assumption of a fixed value of $r$ that bounds the rank of an unknown quantum state from above is necessary, without which it is impossible to decide on the compressive measurement. In real experiments, the value of $r$ requires additional calibrative measurements to be ascertained with a certain degree of credibility, and these are to be counted as part of the resources needed to use such a CS-based procedure. Therefore, the resulting $K_\textsc{ic}$ or $M_\textsc{ic}$ shall only increase. We remark that willfully using a wrong $r$ without a careful verification would generally lead to an untrusted quantum reconstruction. 

Second, any measurement performed in a typical quantum-tomography experiment could incur non-perturbative errors owing to the presence of various noise sources. As such, reliability proofs for CS-based schemes that require noise levels to be perturbative (``$\epsilon$ small''), while correct mathematically, are not applicable to real experiments. An easy way out to circumvent this issue is to compute the fidelity between the reconstructed estimator and a target of interest that one intends to prepare in the first place (assuming a non-adversarial situation). Now, if the fidelity is very high ($>0.98$), then perhaps this may be regarded as a testimony that the reconstruction is indeed faithful with such a scheme. However, it is also usual in practice to acquire a fidelity that is appreciably lower, in which case it is no longer clear whether CS tomography actually works, since without additional treatment, one could never know whether a low fidelity is caused by a set of insufficient measurement outcomes, or due to a distorted rank-$r$ true state---the fidelity is not the correct measure of completeness.

Third, the EP compressive measurements can be difficult to implement either because the POVM outcomes are not rank-one in general or the basis projectors are highly entangled. On the other hand, random Pauli observables, while entirely local, are rather overcomplete by nature, and so it is not unreasonably to perceive a different measurement that can give a smaller $K_\textsc{ic}$. Moreover, owing to the results by Kech and Wolf~\cite{Kech:2016aa} stating that arbitrary von~Neumann bases could be compressive, there is no reason to restrict compressive local measurements to Pauli observables, as arbitrary local bases should also be sufficiently linearly independent to possess a compressive character~\footnote{We shall see in Sec.~\ref{sec:CQST} that it is indeed possible to have a set of local bases that is as compressive as the EP ones (based on numerical evidence), which is much more tomographically efficient than the random Pauli bases.}. A general resource-efficient scheme should thus be able to accommodate such measurements naturally.

\section{Concepts in modern compressive tomography}
\label{sec:modCT}

The main goal of modern compressive tomography is to carry out IC (unique) characterizations of \emph{arbitrary low-rank quantum objects}, be they quantum states, processes, detectors, \emph{etc.}, using minimal measurement resources. Additionally, a practical compressive scheme should satisfy the following criteria:
\begin{enumerate}
	\item No assumptions about the quantum object are needed.
	\item Verifying if a given noisy measurement data is IC must be possible.
	\item Compressive measurements should be feasible to implement in practice.
\end{enumerate}
Demanding that all these criteria be satisfied at once may seem like a tall order, but we shall see that this is, in fact, possible by exploiting intrinsic convex properties of quantum objects.

\subsection{Statistical inference for general quantum tomography}
\label{subsec:stat_inf}

Let us first establish some universal notations. A typical quantum object is specified by $D$ independent parameters that may be consolidated into the column $\rvec{x}$\footnote{Later on, specific tomography applications shall suggest more convenient representations for these parameters. For generalized discussions and pedagogical purposes, these vectorial descriptions are still useful.}. In quantum settings, $\rvec{x}$ should additionally obey a set of convex inequality constraints that can be succinctly written as $\rvec{c}(\rvec{x})\geq\rvec{0}$. This equivalently identifies the convex space $\mathcal{S}$ that contains the admissible parameter column $\rvec{x}$. Measurements are needed to acquire information about an unknown $\rvec{x}$, and these may be encoded into the matrix $\dyadic{A}$. After collecting the measurement data, we may again gather them into the column $\rvec{b}$. When noise is absent, linearity in quantum mechanics implies that $\dyadic{A}\rvec{x}=\rvec{b}$ must hold. Table~\ref{tab:Axb} lists the explicit examples of $\rvec{x}$, $\dyadic{A}$, $\rvec{b}$ and $\mathcal{S}$ in quantum tomography problems.

\begin{table}[t]
	\begin{tabular}{rllll}
		&$\rvec{x}$ & $\rvec{A}$ & $\rvec{b}$ & $\mathcal{S}=\{\rvec{x}|\,\rvec{c}(\rvec{x})\geq\rvec{0}\}$\\\hline\hline\\[-2ex]
		QST: & $\rho$ & POVM & $\tr{\rho\,\Pi_j}$ & $\rho\geq0$, $\tr{\rho}=1$\\
		QPT: & $\Phi$ ($\rho_\Phi$) & States $\{\rho_l\}$, POVM & $\tr{\rho_\Phi\,\rho_l^\top\!\!\otimes\Pi_j}$ & $\rho_\Phi\geq0$, $\mathrm{tr}_{\mathrm{out}}\{\rho_\Phi\}=\openone$\\
		QDT: & $\{\Pi_j\}$ & States $\{\rho_l\}$ & $\tr{\rho_l\Pi_j}$ & $\Pi_j\geq0$, $\sum_j\Pi_j=\openone$\\
		QMT: & $\theta$ & State $\rho_\theta$, POVM & $\tr{\rho_\theta\Pi_j}$ & $0\leq\theta\leq 2\pi$
	\end{tabular}
	\caption{\label{tab:Axb}Common examples of $\dyadic{A}$, $\rvec{x}$, noiseless $\rvec{b}$ and $\mathcal{S}$ in QST~(reconstruction of a state $\rho$), QPT~(reconstruction of the Choi--Jamio{\l}kowski operator $\rho_\Phi$ for a process $\Phi$), QDT~(reconstruction of detector outcomes $\{\Pi_j\}$) and quantum metrology~(QMT; reconstruction of the \emph{classical} interferometric phase $\theta$ with a quantum input state). The entries in the noiseless $\rvec{b}$'s are dictated by the respective versions of Born's rule.}
\end{table}

Real data are noisy, and this would mean that $\dyadic{A}\rvec{x}'=\rvec{b}$ no longer holds in general for all $\rvec{x}'\in\mathcal{S}$. Nevertheless, it is still possible to reconstruct the true parameters $\rvec{x}$ with the help of statistical inference, which provides a map $\mathcal{M}$ that brings $\rvec{b}\mapsto\mathcal{M}[\rvec{b}]=\widehat{\rvec{b}}$\footnote{We denote all \emph{physical} estimators (associated with $\mathcal{S}$) by ``$\widehat{\vphantom{M}\hphantom{M}}$'', which include estimators for $\rvec{x}$ and $\rvec{b}$.}, where $\dyadic{A}\rvec{x}'=\widehat{\rvec{b}}$ is satisfied by some $\rvec{x}'\in\mathcal{S}$. In this article, we shall focus on the following type of inference maps that are commonly used in quantum information:
\begin{definition}[Convex inference maps]
	\label{def:IM}
	\it For a tomography experiment specified by the measurements~$\dyadic{A}$ and (noisy) data~$\rvec{b}$, suppose that a function $g(\dyadic{A}\rvec{x}',\rvec{b})$ is convex in $\rvec{x}'$ and strictly convex in $\dyadic{A}\rvec{x}'$. A convex inference map $\mathcal{M}$ is executed by minimizing $g(\dyadic{A}\rvec{x}',\rvec{b})$ over all $\rvec{x}'\in\mathcal{S}$, such that \mbox{$\dyadic{A}\rvec{x}_\mathrm{min}=\widehat{\rvec{b}}$} is obeyed by the minimum point(s) $\rvec{x}_\mathrm{min}\in\mathcal{S}$. The function $g$ has the following typical characteristics:
	\begin{enumerate}
		\item $\partial g/\partial(\dyadic{A}\rvec{x}_\mathrm{min})=\rvec{0}$ at $\dyadic{A}\rvec{x}_\mathrm{min}=\rvec{b}$ provided that $\rvec{x}_\mathrm{min}\in\mathcal{X}$.
		\item $g$ is additive in the sense that if $\dyadic{A}=(\dyadic{A}_1^\top\,\,\dyadic{A}_2^\top)^\top$ and $\rvec{b}=(\rvec{b}_1^\top\,\,\rvec{b}_2^\top)^\top$ both comprise independent measurements labeled by ``1'' and ``2'', then $g(\dyadic{A}\rvec{x}',\rvec{b})=g(\dyadic{A}_1\rvec{x}',\rvec{b}_1)+g(\dyadic{A}_2\rvec{x}',\rvec{b}_2)$.
	\end{enumerate}
\end{definition}
\noindent
A simple consequence of using such a convex inference map is that the resulting physical $\widehat{\rvec{b}}$ is \emph{always} unique regardless of whether $\rvec{x}_\mathrm{min}$ is or not, since it is equivalent to the minimization of $g(\widehat{\rvec{b}},\rvec{b})$, which is a strictly convex function of $\widehat{\rvec{b}}$, over a convex~$\mathcal{C}$. The additivity characteristic endows $g$ with an information-like property.

It is sensible to process data into a column $\rvec{b}$ that is \emph{statistically consistent}, that is $\rvec{b}$ and its physical counterpart $\widehat{\rvec{b}}$ both approach the true data given by $\rvec{x}$ for sufficiently large data sample size. On this premise, it is desirable for convex inference methods to also give a unique estimator $\widehat{\rvec{x}}\rightarrow\rvec{x}$ whenever $\dyadic{A},\rvec{b}$ is IC. Indeed, this follows from characteristics~(1) in {\bf Definition~\ref{def:IM}}, so long as $\rvec{b}$ is statistically consistent. Popular methods like the ML~\cite{Aldrich:1997ml,Banaszek:1999ml,Rehacek:2007ml,Teo:2011me,Teo:2012cs,Teo:2012ve,Shang:2017sf} and least-squares~(LS)~\cite{Opatrny:1997aa,James:2001aa,Langford:2007th,Yuanlong:2019} procedures fall under the category of convex inference maps in regular physics settings. 

As concrete examples, in QST ($D=d^2-1$), it is convenient to introduce a set of Hermitian traceless basis operators $\{\Omega_l\}^{d^2-1}_{l=1}$, where $\tr{\Omega_{l}\Omega_{l'}}=\delta_{l,l'}$. In terms of this operator basis, the true parameter $\rvec{x}$ is the column of $d^2-1$ quantum parameters for a $d$-dimensional state $\rho=\openone/d+\sum^{d^2-1}_{l=1}x_l\,\Omega_l$. Given an $M$-outcome POVM $\sum^M_{j=1}\Pi_j=\openone$ that is used to probe $\rho$, with which the relative frequencies $\sum^M_{j=1}\nu_j=1$ are obtained, we have
\begin{equation}
	\dyadic{A}=\begin{pmatrix}
		\tr{\Pi_1\Omega_1} & \tr{\Pi_1\Omega_2} & \cdots & \tr{\Pi_1\Omega_{d^2-1}}\\
		\tr{\Pi_2\Omega_1} & \tr{\Pi_2\Omega_2} & \cdots & \tr{\Pi_2\Omega_{d^2-1}}\\
		\vdots & \vdots & \ddots & \vdots\\
		\tr{\Pi_M\Omega_1} & \tr{\Pi_M\Omega_2} & \cdots & \tr{\Pi_M\Omega_{d^2-1}}
	\end{pmatrix}\,,\,\,
	\rvec{b}=\begin{pmatrix}
		\nu_1-\tr{\Pi_1}/d\\
		\nu_2-\tr{\Pi_2}/d\\
		\vdots\\
		\nu_M-\tr{\Pi_M}/d
	\end{pmatrix}\,,
	\label{eq:Ab_state}
\end{equation}
where $\rvec{b}$ is statistically consistent because $\nu_j\rightarrow p_j=\tr{\rho\Pi_j}$, and $\dyadic{A}\rvec{x}=\rvec{p}-\rvec{\Pi}/d$ with $\rvec{\Pi}$ being the column ($\tr{\Pi_l}$). The LS inference method then uses the convex function $g$ defined as
\begin{equation}
	g=g_\textsc{ls}(\dyadic{A}\rvec{x}',\rvec{b})\equiv\|\dyadic{A}\rvec{x}'-\rvec{b}\|^2\,,
	\label{eq:LS}
\end{equation}
where characteristics~(1) and (2) evidently hold. On the other hand, the ML inference method requires the correct form of the likelihood function $L(\rvec{x}'|\rvec{\nu})=L(\rho'|\rvec{\nu})$. For instance, in a usual multishot tomography experiment where each detector count is received independently until a \emph{fixed} total number of counts ($N$) is acquired, the \emph{log-likelihood} $\log L(\rho'|\rvec{\nu})=N\sum^M_{j=1}\nu_j\log \tr{\rho'\Pi_j}$ takes a multinomial form. The relevant convex function 
\begin{align}
	g=g_\textsc{ml}(\dyadic{A}\rvec{x}',\rvec{b})\equiv&\,-N\sum^M_{j=1}\left(\rvec{b}+\dfrac{\rvec{\Pi}}{d}\right)_j\log\left(\dyadic{A}\rvec{x}'+\dfrac{\rvec{\Pi}}{d}\right)_j\nonumber\\
	=&\,-N\sum^M_{j=1}\nu_j\log\tr{\rho'\Pi_j}
	\label{eq:ML}
\end{align}
would again possess all the required characteristics. In particular, the minimum value of $g_\textsc{ml}$ is stationary provided that there exists some $\rho'$ belonging to the quantum state space $\mathcal{S}$ for which $\tr{\rho'\Pi_j}=\nu_j$ ($\dyadic{A}\rvec{x}'=\rvec{b}$). 

We note that there is yet the \emph{Bayesian mean}~(BM)~\cite{Blume-Kohout:2006aa,Blume_Kohout_2010} inference method, with which the state estimator
\begin{equation}
	\widehat{\rho}_\textsc{bm}=\dfrac{\int_\mathcal{S}(\D\rho')L(\rho'|\rvec{\nu})\,\rho'}{\int_\mathcal{S} (\D\rho')L(\rho'|\rvec{\nu})}
\end{equation}  
is defined as the state-space average weighted according to the likelihood $L(\rho'|\rvec{\nu})$ and volume measure $(\D\rho')$ that contains an assigned prior distribution necessary in such a Bayesian analysis. In the limit of large $N$, the posterior measure $(\D\rho')L$ becomes sharply peaked around the maximum of $L$, so that BM essentially becomes ML in this asymptotic limit. Presented with a zoo of statistical inference methods, we stress that a good choice should be guided by common sense in relation to the actual physical situation.

\subsection{Compression with convex quantum constraints}
\label{subsec:comp_cvx}

Suppose that we would like to recover a $D$-dimensional $\rvec{x}'$ that reconstructs some unknown $\rvec{x}$, and $\dyadic{A}$ is a $M\times D$ matrix. Then, basic linear algebra tells us that the linear system $\dyadic{A}\rvec{x}'=\widehat{\rvec{b}}$ has no unique solution $\rvec{x}'$ if $M<D$. In other words, the solution space $\mathcal{X}$ for $\rvec{x}'$ corresponding to this linear system will always be infinite unless $M\geq D$. The motivation of compressive tomography, however, stems from the recognition that $\rvec{x}'$ must additionally belong to the physical space $\mathcal{S}$, and so $\rvec{x}'\in\mathcal{X}\cap\mathcal{S}\subset\mathcal{X}$. The resulting physical search space is thus much smaller than the unconstrained solution space $\mathcal{X}$. 

\begin{figure}[t]
	\centering
	\includegraphics[width=0.5\columnwidth]{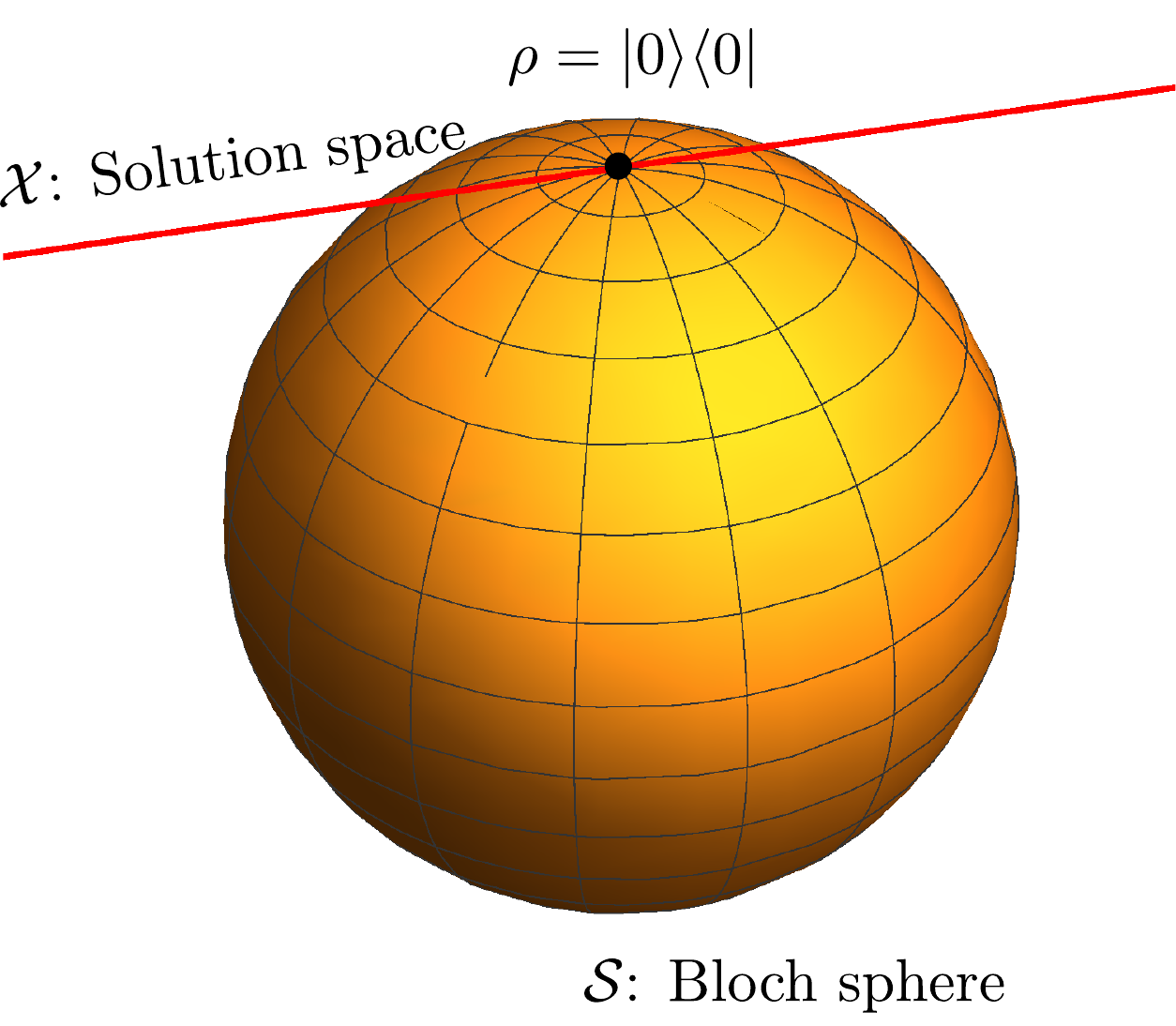}
	\caption{\label{fig:bloch}Single-qubit tomography compression with the Bloch sphere. In this case, $\mathcal{X}\cap\mathcal{S}=\{\rho=\ket{0}\bra{0}\}$ at $M=2<D$.}
\end{figure}

As an instructive example, we look at single-qubit QST~($D=3$). Suppose the expectation values of two projectors, $\ket{\phi_1}\bra{\phi_1}=[\openone+(-\sigma_x+\sigma_y+\sigma_z)/\sqrt{3}]/2$ and $\ket{\phi_2}\bra{\phi_2}=[\openone+(\sigma_x-\sigma_y+\sigma_z)/\sqrt{3}]/2$, for the unknown qubit quantum state $\rho$ are measured, and we obtain the corresponding relative frequencies $\nu_1=(1+1/\sqrt{3})/\sqrt{2}=\nu_2$. Upon expressing the $2\times3$ matrix 
\begin{equation}
	\dyadic{A}=\begin{pmatrix}
		\opinner{\phi_1}{\sigma_x}{\phi_1} & \opinner{\phi_1}{\sigma_y}{\phi_1} & \opinner{\phi_1}{\sigma_z}{\phi_1}\\
		\opinner{\phi_2}{\sigma_x}{\phi_2} & \opinner{\phi_2}{\sigma_y}{\phi_2} & \opinner{\phi_2}{\sigma_z}{\phi_2}
	\end{pmatrix}
\end{equation}
in terms of these two rank-one projector measurements and defining $\rvec{b}=2(\nu_1\,\,\,\nu_2)^\top-1=(1\,\,\,1)^\top/\sqrt{3}$\footnote{The definitions of $\dyadic{A}$ and $\rvec{b}$ slightly deviate from \eqref{eq:Ab_state} to go in accordance with the standard Bloch representation.}, it is obvious that $M=2<D$ gives an infinite line as the solution space---$\mathcal{X}=\{(x\,\,y\,\,z)^\top|\,y=x\,\&\,z=1\}$. In this case, there exists an estimator $\widehat{\rvec{x}}$ in $\mathcal{S}$, which is the Bloch sphere, that satisfies $\dyadic{A}\widehat{\rvec{x}}=\rvec{b}$, and the intersection $\mathcal{X}\cap\mathcal{S}$ contains precisely a single point, namely the state $\rho=(\openone+\sigma_z)/2=\ket{0}\bra{0}$~(see Fig.~\ref{fig:bloch}). So, with just two projectors and the Bloch-sphere constraint, we are able to uniquely characterize all three parameters in $\rvec{x}$.

Very generally, for an unknown low-rank quantum object parametrized by $\rvec{x}$ that is probed by a collection of measurements encoded in $\dyadic{A}$, the objective of compressive tomography is to build up an $\dyadic{A}$ for which $\mathcal{C}\equiv\mathcal{X}\cap\mathcal{S}\ssubset\mathcal{X}$ happens quickly, such that a unique estimator $\widehat{\rvec{x}}$ is obtained for just $M\ll D$ measurements. For sufficiently large data samples, any statistically consistent inference mapping would result in $\widehat{\rvec{x}}\rightarrow\rvec{x}$. This leads to the question of how one can validate whether the convex search space $\mathcal{C}$ is a point or not. The answer to this question represents the functional core of compressive tomography, which is addressed in the next section.  

\subsection{Informational completeness certification}
\label{subsec:ICC}

We suppose the given measurement outcomes $\dyadic{A}$, their collected data $\rvec{b}$ and corresponding physical $\widehat{\rvec{b}}$ acquired using a statistically-consistent convex inference method described in {\bf Definition~\ref{def:IM}}. The convex quantum constraints for $\rvec{x}$ results in the convex subspace $\mathcal{C}=\mathcal{X}\cap\mathcal{S}$ that contains all possible estimators $\rvec{x}'$ that are consistent with $\widehat{\rvec{b}}$. The task of \emph{informational completeness certification}~(ICC) is to correctly determine whether $\mathcal{C}$ is a point or not. 

Numerically, ICC is equivalent to an optimization problem; namely, if we define $f(\rvec{x}')$ to be a \emph{strictly} concave\footnote{For $0\leq\lambda\leq1$, $f(\lambda\rvec{x}'_1+(1-\lambda)\rvec{x}'_2)>\lambda f(\rvec{x}'_1)+(1-\lambda)f(\rvec{x}'_2)$. A strictly convex $f$ also works.} function of $\rvec{x}'\in\mathcal{C}$, then the minimum $f_\mathrm{min}$ and maximum $f_\mathrm{max}$ of $f$ over $\mathcal{C}$ can be used to find out if $\mathcal{C}$ has a finite volume. More specifically, the quantity $s_\textsc{cvx}\propto f_\mathrm{max}-f_\mathrm{min}$ is an IC~indicator that possesses the following important properties (arguments supplied in~\ref{app:prop}):

\begin{prop}[Noiseless monotonicity]
	\label{prop:monotone}
	\it For noiseless data, suppose that the pairs $\dyadic{A}'$,$\rvec{b}'$ and $\dyadic{A}$,$\rvec{b}$ are such that the former exactly includes the latter. Then $s'_{\textsc{cvx}}\leq s_{\textsc{cvx}}$.  
\end{prop}

\begin{prop}[IC]
	\label{prop:IC}
	\it For a strictly-concave $f$, $s_\textsc{cvx}=0$ implies that $\mathcal{C}$ contains a single point for \emph{any} $\rvec{b}$.
\end{prop}

\begin{prop}[Noisy IC]
	\label{prop:trueIC}
	\it Suppose that $\dyadic{A}=(\dyadic{A}_1^\top\,\,\dyadic{A}_2^\top)^\top$ contains two sets of sequentially measured outcomes, first of $\dyadic{A}_1$ and next of $\dyadic{A}_2$, which are accompanied by (noisy) data $\rvec{b}=(\rvec{b}_1^\top\,\,\rvec{b}_2^\top)^\top$. Then any statistically-consistent convex inference map outlined in {\bf Definition~\ref{def:IM}} implies that if $s^{(1)}_{\textsc{cvx}}=0$, it must also be that $s^{(1,2)}_{\textsc{cvx}}=0$ provided that characteristic~(1) is fulfilled for $\dyadic{A}$ and $\rvec{b}$.
\end{prop}
\noindent

The validity of the noisy IC property ({\bf Property~\ref{prop:trueIC}}) hinges on the \emph{existence} of an $\rvec{x}_\mathrm{min}$ lying in the general solution space $\mathcal{X}$ such that $\dyadic{A}\rvec{x}_\mathrm{min}=\rvec{b}$, so that convexity in $g$ that is used by the convex inference map implies that $\mathcal{C}$ is the set of states consistent with $\rvec{b}$. This condition is typically violated for any noisy $\rvec{b}$ so long as $\dyadic{A}$ has sufficiently many rows. This, of course, makes an elementary connection with basic linear algebra: a noisy linear system of equations would almost surely conflict with one another to give \emph{no} solution, let alone a physical one.\footnote{In QST, for example, when the POVM is a small set of generic von~Neumann bases, such a conflict almost never happens if $M<D$. However, for \emph{local} (product) bases, this happens rather often because a non-IC set of such bases can extract linearly-dependent noisy information~(see Sec.~\ref{sec:CQST_generic}).} Thus, a scenario in which $s^{(1,2)}_\textsc{cvx}>s^{(1)}_\textsc{cvx}=0$ ($\mathcal{C}^{(1,2)}>\mathcal{C}^{(1)}$) can arise in practice, and such is not a result of any specific defect in modern compressive tomography or previous CS-related schemes, but rather an inevitable result of $\mathcal{C}$'s statistical fluctuation. Another crucial remark is that $s^{(1,2)}_\textsc{cvx}>0$ does not, in any way, mean that the previous IC result $s^{(1)}_\textsc{cvx}=0$ is wrong, as $\mathcal{C}^{(1,2)}$ and $\mathcal{C}^{(1)}$ are, respectively, two separate convex sets of states associated with different physical $\widehat{\rvec{b}}^{(1,2)}\neq \widehat{\rvec{b}}^{(1)}$. A silver lining here is that such variations of $s_\textsc{cvx}$ asymptotically vanish for sufficiently large data samples for statistically consistent inference maps, which is anyway needed for accurate tomography, and {\bf Property~\ref{prop:trueIC}} asymptotically holds even in such cases. This means that in a bottom-up compressive scheme where measurements are performed one after another, extracting the unique estimator at the first instance when $s_\textsc{cvx}=0$ is still a statistically reasonable option.

On a related note, we mention in passing that the effect of informational incompleteness is masked by the BM inference method discussed in Sec.~\ref{subsec:stat_inf}, since $\widehat{\rho}_\textsc{bm}$ is always a unique full-rank state estimator. 

Based on these properties, we may concisely state the simple numerical procedure of ICC as {\bf Algorithm~\ref{algo:ICC}} in~\ref{app:algos}. Harking back to QST as our exemplifying candidate, a strictly concave $f$ can be defined by $f(\rho')=\tr{\rho'Z}$ in terms of the variable quantum state $\rho'$ and full-rank state $Z$ that can be randomly chosen, but otherwise fixed. This is the safest option to avoid pathological cases. (a)~For instance, $Z=1/d$ is obviously forbidden for the trivial reason that $f(\rho')=1$ for all $\rho'$. (b)~Next, if we take $Z=\ket{\phi}\bra{\phi}$ to be some pure state, then any $\rho'$ that lies in the kernel of $Z$ shall always give $s_\textsc{cvx}=0$. (c)~Now, let us consider all states $\rho'\in\mathcal{C}$ such that $\opinner{j'}{\rho'}{j'}=\widehat{p}_{j'}$ with respect to a single complete basis $\mathcal{B}=\{\ket{j'}\}$ as the POVM and a full-rank $Z$. If $\mathcal{B}$ is the eigenbasis of $Z=\sum^{d-1}_{j'=0}\ket{j'}z_{j'}\bra{j'}$, then clearly $f(\rho')=\sum^{d-1}_{j'=0}\widehat{p}_{j'}z_{j'}=\mathrm{constant}$ for this entire $\mathcal{C}$, which illustrates yet another serious pathological situation that wrongly suggests that a single basis measurement is IC. So, if $Z$ is a randomly-chosen full-rank state, then such special cases are measure zero, and we always have $f_\mathrm{max}>f_\mathrm{min}$ unless the measurements are IC.

The ICC algorithm is therefore a semidefinite program~(SDP) since a linear function is optimized within the constraint subspace $\mathcal{C}$ that is convex. This implies that the computational complexity in determining whether a given measurement set is IC or not turns out to be polynomial in the dimension of the parameter space $D$~\cite{Vandenberghe:1996ca}, and requires no assumptions whatsoever about the unknown true $\rvec{x}$. This is completely different from ascertaining whether a set of measurements possess the RIP property, which is NP-hard~\cite{Tillmann:2006aa}.

\section{Compressive quantum state tomography}
\label{sec:CQST}

From hereon, unless otherwise stated, we shall consider the multinomial setting and invoke the ML procedure in accordance with Eq.~\eqref{eq:ML} as the default inference map~\footnote{This can be numerically carried out either with the steepest-ascent gradient~\cite{Rehacek:2007ml} or the accelerated projected-gradient method~\cite{Shang:2017sf}, where MATLAB codes concerning the latter for QST are available at \url{https://github.com/qMLE/qMLE}.}. Accordingly, in the language of QST, the measurement probing a given unknown $d$-dimensional state $\rho$ are POVM outcomes with their corresponding relative frequencies. In compressive QST~(CQST), the task is to reconstruct some unknown rank-$r$ state $\rho=\rho_r$ with no \emph{a priori} assumptions about $\rho$\footnote{A realistically generated $\rho$ is full-rank. However, CQST still works well if $\rho$ has a rapidly decreasing eigenspectrum, which is the situation for well-controlled sources necessary for quantum tasks.}. The ICC routine in {\bf Algorithm~\ref{algo:ICC}} needed for CQST is now governed by the linear function $f=\tr{\rho'Z}$ specified by a fixed full-rank state $Z$ that is randomly chosen for reasons discussed in Sec.~\ref{subsec:ICC}. The quantum constraints $\rho'\geq0$ and $\tr{\rho'}=1$ are already in their native forms that are ideal for executing an SDP.

In practice, von~Neumann measurement bases are a favorite choice of POVM as each basis may be measured independently in a bottom-up manner to build a small measurement set that is just sufficient for an IC state reconstruction. Notation-wise, we then have a set of $K$ measured von~Neumann bases, where the $k$th basis $\mathcal{B}_k=\{\ket{b_{jk}}\}^{d-1}_{j=0}$ gives the relative frequencies $\rvec{\nu}_k=\{\nu_{jk}\}^{d-1}_{j=0}$. For $1\leq k\leq K$, these are properly normalized as $\sum^{d-1}_{j=0}\ket{b_{jk}}\bra{b_{jk}}=1$ and $\sum^{d-1}_{j=0}\nu_{jk}=1$. The relative frequencies asymptotically converge to the true probabilities $p_{jk}=\opinner{b_{jk}}{\rho}{b_{jk}}$ in the limit of large number of copies $N$, which is now defined \emph{per measured basis}. Otherwise, any finite value of $N$ would result in statistically noisy $\nu_{jk}$, from which ML converts to the physical probabilities $\widehat{p}_{jk}$.

\subsection{Generic-bases tomography}
\label{sec:CQST_generic}

A natural first question to ask is whether a set of randomly generated entangled von~Neumann bases $\mathcal{B}=\{\mathcal{B}_1,\mathcal{B}_2,\ldots,\mathcal{B}_K\}$ can be compressive. The intuitive answer is a positive one since any arbitrary set of $d+1$ bases form the minimal POVM needed to reconstruct any strictly full-rank $d$-dimensional quantum state with no rapidly decreasing eigenspectrum. Indeed, if $\rho=\rho_r$, the work of Kech and Wolf~\cite{Kech:2016aa}~(KW) supports this answer by giving the sufficient IC requirement $K>K_\textsc{kw}\equiv[4r(d-r)-2]/(d-1)$ for $r\leq d/2$, which is generally a loose upper bound of $K_\textsc{ic}$ for arbitrary $d$. The goal is then to determine the exact value of $K_\textsc{ic}$ for every set of random bases and data.

For more feasible measurements, it is much more desirable to implement local von~Neumann bases. For a given $n$-partite state of $d=d_0^n$, these are defined by $\mathcal{B}^{\text{loc}}_k=\{\ket{b^{(1)}_{j_1k}}\otimes\cdots\otimes\ket{b^{(n)}_{j_nk}}\}^{d_0-1}_{j_m=0}$, where $\sum^{d_0-1}_{j_m=0}\ket{b^{(m)}_{j_mk}}\bra{b^{(m)}_{j_mk}}$ for all $1\leq m\leq n$. A classic example pertains to $n$-qubit systems~($d_0=2$), for which the overcomplete set $\mathcal{B}^{\text{loc}}$ comprises $3^n$ local von~Neumann bases that may be chosen as the eigenbases of Pauli operators. To characterize any strictly full-rank $n$-partite qudit state, one needs a minimum of $(d_0+1)^n$ local bases, where each subsystem are uniquely specified with $d_0+1$ independent constraints, which is much larger than $d_0^n+1$ for $n\gg1$. Therefore, a bottom-up CQST scheme for low-rank states that pertains solely to local-bases measurements is definitely in order.

It is easy to build a bottom-up iterative CQST scheme for both kinds of random basis measurements. We can start with the standard computational basis $\mathcal{B}_1=\{\ket{j'}\bra{j'}\}^{d-1}_{j'=0}$, where $\ket{j'}\,\widehat{=}\,(0,0,\ldots,1,0,\ldots,0)^\top$ is represented by a column of zeros except at the $j'$th position where the entry is~1. For multipartite systems, $\mathcal{B}_1$ is clearly a product basis. After measuring $\mathcal{B}_1$, the corresponding data $\rvec{\nu}_1$ is processed with the ML procedure to obtain $\widehat{\rvec{p}}_{\textsc{ml},1}$, and ICC is then carried out to compute $s_{\textsc{cvx},1}$. Obviously, $s_{\textsc{cvx},1}$ cannot be zero at this point since $\mathcal{B}_1$ simply identifies the diagonal elements of $\rho_r$ represented in this basis. So, a second basis $\mathcal{B}_2$, either entangled or local, is chosen for the measurement and the combined data $\rvec{\nu}=(\rvec{\nu}_1^\top\,\,\,\rvec{\nu}_2^\top)^\top$ are again subjected to ML processing and ICC computation is carried out to yield $s_{\textsc{cvx},2}$. This iterative scheme continues until step~$K=K_\textsc{ic}$ where $s_{\textsc{cvx},K_\textsc{ic}}=0$ (see {\bf Algorithm~\ref{algo:HaarCQST}}).

\begin{figure}[t]
	\centering
	\includegraphics[width=0.8\columnwidth]{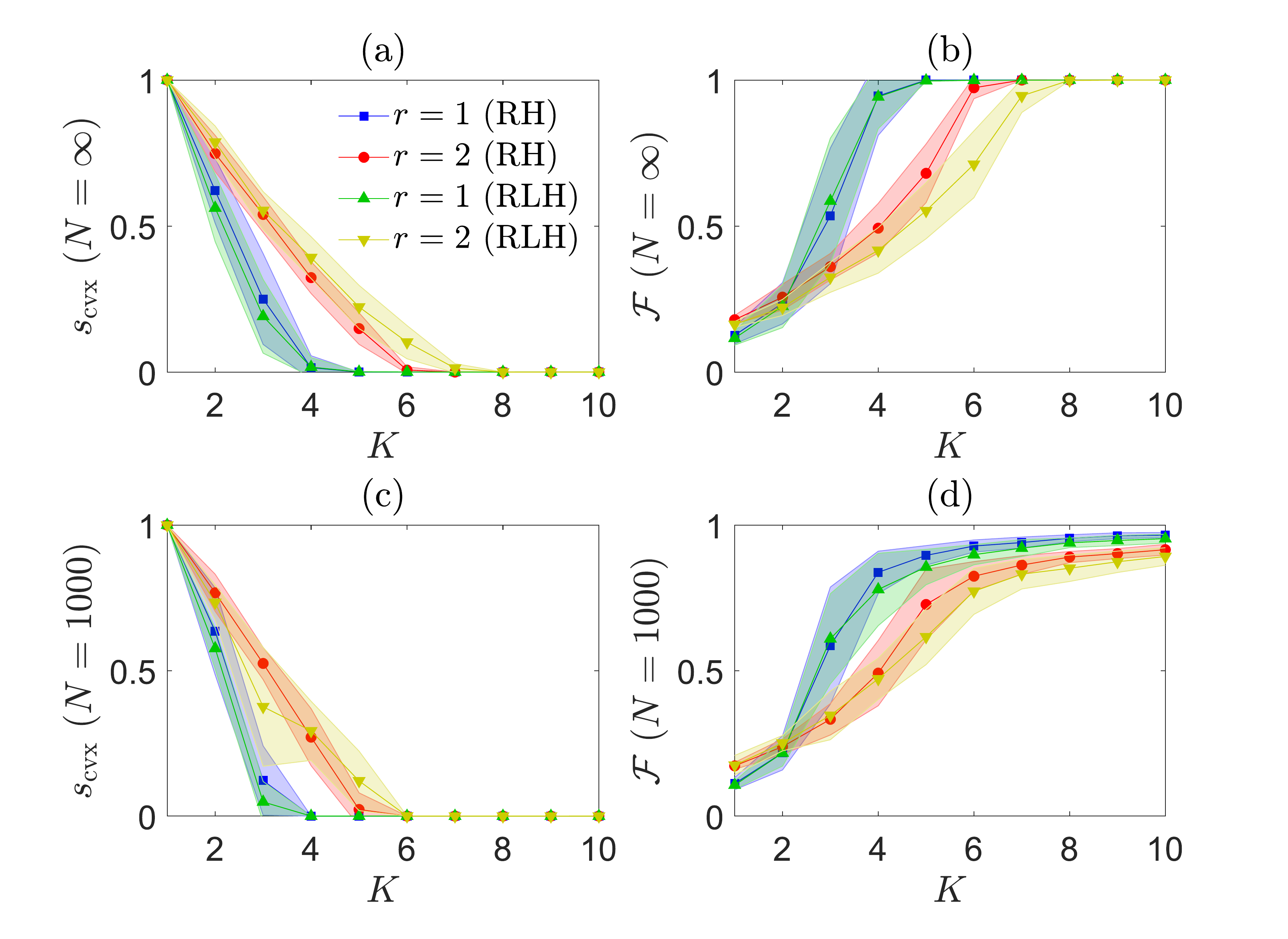}
	\caption{\label{fig:Haard16}A showcase of how the characteristics of both the $s_\textsc{cvx}$ and $\mathcal{F}$ curves typically look like with respect to $K$. As an example, we plot these curves for random-bases CQST on $d=16$ systems. Both entangled~(ent) and local~(loc, $d_0=2$) bases are considered. All fidelity curves in (b) and (d) are plotted with the ML state estimator obtained using $1/d$ as the seed state. Local bases require a larger $K_\textsc{ic}$ value to be IC due to their overcomplete nature. All $s_\textsc{cvx}$ values are normalized to 1, and the 1-$\sigma$ error regions refer to the statistics of 10 randomly generated states for each rank $r$.}
\end{figure}

We look at random bases distributed according to the Haar measure, which is an extremely popular choice in quantum information theory~\cite{Cwiklinski:2013aa,Russell:2017aa,Banchi:2018aa}. It is known that a variant of the QR decomposition generates unitary operators distributed according to this measure~\cite{Mezzadri:2007qr}, so that {\bf Algorithm~\ref{algo:HaarB}} applies. Another routine that we shall consistently rely on is the generation of random rank-$r$ states for simulating the results in this article. The corresponding {\bf Algorithm~\ref{algo:randstate}} highlights a simple recipe to generate random states distributed uniformly according to the Hilbert--Schmidt measure~\cite{Zyczkowski:2003hs}. Figure~\ref{fig:Haard16} shows the general behaviors of $s_\textsc{cvx}$ and reconstruction fidelity $\mathcal{F}$ with the number of measured bases~$K$ for four-qubit systems~($d=2^4=16$) under the random Haar~(RH) and random local Haar~(RLH) bases schemes.

\subsection{Adaptive-bases tomography}

\begin{figure}[t]
	\centering
	\includegraphics[width=1\columnwidth]{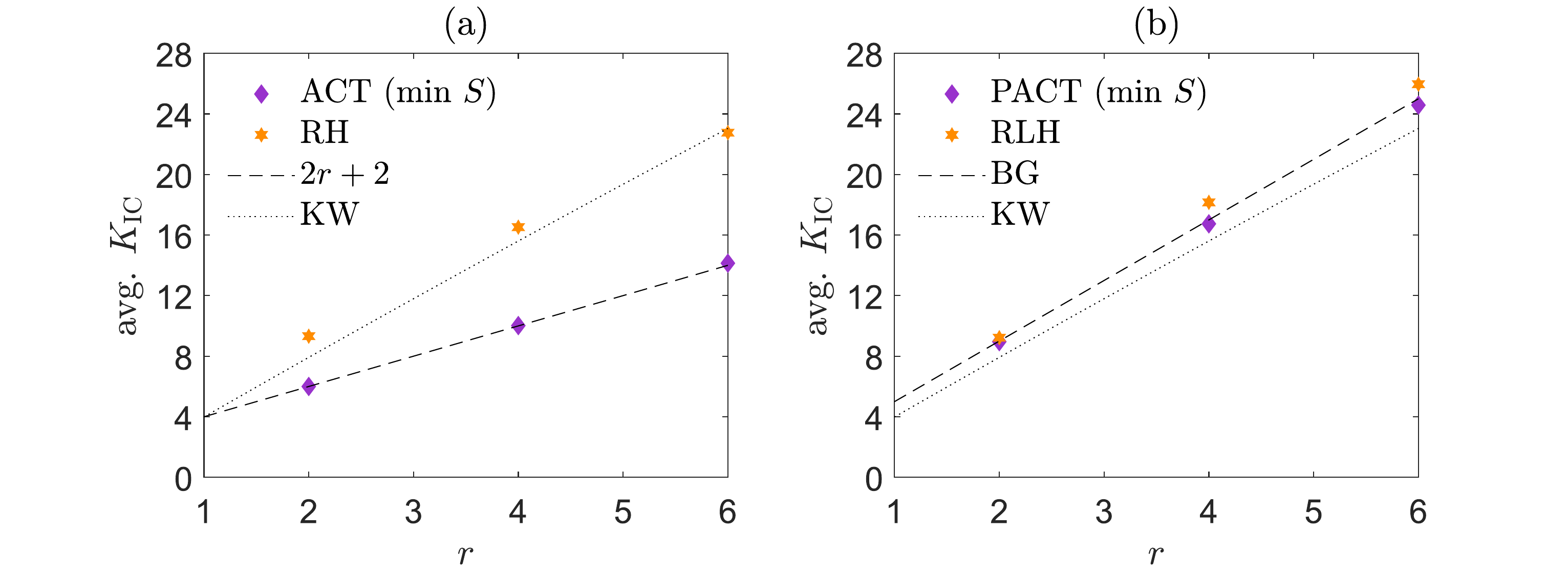}
	\caption{\label{fig:comp_CQST}A noiseless comparison of $K_\textsc{ic}$ (averaged over 50 states per marker) with seven-qubit systems $(d=2^7=128)$ for all key protocols and known scaling behaviors for benchmarking purposes. In this regime of large $d$, the $K_\textsc{ic}$ scaling for ACT converges to $2r+2$, and that for PACT to $4r+1$, which is the same scaling behavior of the BG scheme. For such a moderately large system, the performance gap between ACT and RH is clear, whereas optimizations over the local-unitary space seem to introduce limited improvement in comparison to the RLH scheme. We see that the $K_\textsc{ic}$ for both random schemes are higher than the KW benchmark as they should. For the adaptive schemes, ACT can significantly break the KW barrier. PACT, on the other hand, is almost as good as RLH.}
\end{figure}

The next progressive step would be to establish an adaptive procedure in efforts to reduce the $K_\textsc{ic}$ value. To see how one can do this straightforwardly, we give a precise statement of state-space-boundary-induced compression for von~Neumann-bases QST that is relevant for constructing such an adaptive procedure:

\vspace{2ex}
\noindent
{\bf State-space-boundary-induced compression:} {\it Suppose statistical fluctuation is absent, and that it so happens the basis $\mathcal{B}_{\rho_r}$ that contains all support eigenkets of $\rho=\rho_r$, unbeknownst to the observer, is measured. If one measures \mbox{$K_0=1+\lceil(r^2-r)/(D-1)\rceil$} orthonormal bases that include $\mathcal{B}_{\rho_r}$, then as a consequence of the state-space boundary, $\widehat{\rho}=\rho_r$ is the unique positive estimator consistent with all measured probabilities.}
\vspace{2ex}

We can understand the validity of this statement through the inspection of easy special cases. For pure states $(r=1)$, measuring the basis containing $\rho_1$ is equivalent to knowing the state completely; only the projector $\rho_1$ in this basis would ``click''. Hence, $K_0=1$. If $r=D$, we yet obtain the familiar result $K_0=d+1$. While the justification is furnished in~\ref{app:bd_thm}, we note that when $r^2-r\leq d-1$, measuring just one basis other than $\mathcal{B}_{\rho_r}$ will result in a unique $\widehat{\rho}=\rho_r$, which exhibits strong compression ($K_0=2$). Obviously, since we posit that $\rho_r$ is completely unknown, the above statement cannot be applied directly. Thus, the famous no-go answer $K_\textsc{ic}>2$ to the Pauli  problem~\cite{Pauli:1933pr,Carmeli:2015cs} of arbitrary pure-state reconstruction is never invalidated. Nonetheless, it still expresses a crucial notion of compressive tomography---the state-space boundary induces a reduction in $K_\textsc{ic}$, which is consistent with results of an earlier work~\cite{Kalev:2015aa}. 

Very favorably, the manifestation of compression through the state-space boundary offers a useful hint for establishing \emph{adaptive compressive tomography}~(ACT); it suggests (in the noiseless) the possibility that even for an unknown $\rho_r$, one can eventually find out $\mathcal{B}_{\rho_r}$ with more bases measurements, at which point tomography can be reliably terminated since $K_\textsc{ic}>K_0$. It turns out that ACT can be formulated in an iterative scheme that applies to arbitrary noisy datasets. In every iteration step, the trick is to cleverly choose the next optimal measurement basis that can quickly converge to $\mathcal{B}_{\rho_r}$. To this end, we recognize that compressive tomography works best for $r\ll d$, and that the logical optimal basis to choose is the eigenbasis of some rank-deficient state in $\partial \mathcal{S}\cap\mathcal{C}$. We know that the minimum-von~Neumann entropy~(minENT, $S(\rho')=-\tr{\rho'\log\rho'}$) estimator over $\mathcal{C}$ is a good candidate, since entropy minimization has rank-reducing tendencies and optimizing $S$ over $\mathcal{C}$ guarantees convergence to a statistically-consistent unique estimator as $K$ increases, and has also been suggested as an efficient method in conventional compressed sensing applications~\cite{Tran:2016aa}. Taking the minENT approach, we introduce the ACT~\cite{Ahn:2019aa,Ahn:2019ns} scheme in {\bf Algorithm~\ref{algo:ACT}} for adaptively reducing $K_\textsc{ic}$. The local version of ACT~(PACT) picks the next local basis closest to the eigenbasis of the minENT estimator. One option might be to minimize some distance measure between the unitary operator $U_\text{minENT}$ that diagonalizes the minENT estimator and $U'_1\otimes\cdots\otimes U'_n$ over all subsystem-unitary spaces. 

It has been conjectured with numerical evidence~\cite{Ahn:2019ns} that in the limit of large $d$, we have $K_\textsc{ic}=2r+2$ for ACT and $K_\textsc{ic}=4r+1$ for PACT at least in the regime $r\ll d$ (tested up to $r=6$). Based on these scalings, one needs about twice as many IC local bases as IC entangled bases on average for reconstructing an unknown state of a large rank. Interestingly, the latter scaling exactly coincides with the BG scheme, although the average $K_\textsc{ic}$ gap between PACT and RLH is generally not large in contrast with that between ACT and RH. The lesson here is that one can achieve the scaling of a set of entangled bases feasibly using only adaptive local bases, which is encouraging as this means that CQST can be carried out on NISQ devices. We note, however, that this does not discount the value of entangled BG bases~\cite{Baldwin:2016cs}, for they are explicit constructions that gave the $K_\textsc{ic}$ value for \emph{all states of rank no larger than some known integer $r$}. While adaptive graphs similar to those in Fig.~\ref{fig:Haard16} may be plotted for $d=16$, Fig.~\ref{fig:comp_CQST} presents a comparison of all the aforementioned CQST schemes for seven-qubit systems~($d=2^7=128$), which has a sufficiently large Hilbert-space dimension on which the conjectured scaling behaviors are based.

\section{Compressive quantum process tomography}
\label{sec:CQPT}

We now turn to compressive quantum process tomography~(CQPT), which is the compressive reconstruction of an unknown low-rank quantum channel or process $\Phi$ that maps a quantum state to another: $\Phi[\rho_\textsc{in}]=\rho_\textsc{out}$\footnote{Here, we always consider cases where $\mathrm{dim}\{\rho_\textsc{in}\}=d=\mathrm{dim}\{\rho_\textsc{out}\}$, but generalization to arbitrary input and output dimensions is straightforward.}. A complete knowledge of $\Phi$ therefore implies that one should be able to characterize $\Phi$ that is completely-positive~(CP) and trace-preserving~(TP). For $d$-dimensional systems, one natural way of representing $\Phi$ is to send a subsystem through it from an entangled two-qudit system prepared in a bipartite maximally-entangled state $\ket{\textsc{me}}\bra{\textsc{me}}$, where $\ket{\textsc{me}}=\sum^{d-1}_{j=0}\ket{j}_1\ket{j}_2/\sqrt{d}$ for the computational basis. The resulting (unnormalized) output state is then the so-called Choi--Jamio{\l}kowski~\cite{Choi:1975aa,Jamiolkowski:1972aa} operator representation of $\Phi$, that is $\rvec{x}\leftrightarrow\rho_\Phi\equiv\mathcal{I}\otimes\Phi[\ket{\textsc{me}}d\bra{\textsc{me}}]$, where $\mathcal{I}$ is the identity map or idle process. The action of $\Phi$ can then be expressed in terms of $\rho_\Phi$ as $\rho_\textsc{out}=\ptr{1}{\rho_\Phi\,\rho_\textsc{in}^\top\otimes \openone}$~\footnote{The transposition is always referring to the computational basis.}. This follows swiftly from the computational-basis representation $\rho_\textsc{in}=\sum^{d-1}_{l,l'=0}\ket{l}\rho^\textsc{in}_{l,l'}\bra{l'}$,
\begin{align}
	\ptr{1}{\rho_\Phi\,\rho_\textsc{in}^\top\otimes 1}=&\,\ptr{1}{\sum^{d-1}_{j=0}\sum^{d-1}_{j'=0}\sum^{d-1}_{l=0}\ket{j}\rho^\textsc{in}_{lj'}\bra{l}\otimes\Phi[\ket{j}\bra{j'}]}\nonumber\\
	=&\,\sum^{d-1}_{j=0}\sum^{d-1}_{j'=0}\rho^\textsc{in}_{jj'}\Phi[\ket{j}\bra{j'}]\nonumber\\
	=&\,\Phi[\rho_\textsc{in}]=\rho_\textsc{out}\,.
\end{align}
Apart from being a positive operator that originates from representing a physical CP process, $\rho_\Phi$ should also possess the additional TP constraint, which is equivalent to the requirement that $\tr{\rho_\textsc{out}}=1$ holds for \emph{all} $\rho_\textsc{out}$, and straightforwardly implies that 
\begin{equation}
	\ptr{2}{\rho_\Phi}=\openone\,.
\end{equation}
Clearly, $\rho_\Phi$ already contains $d^4$ real parameters due to Hermiticity, so the TP constraint reduces the total number of parameters to $D=d^4-d^2$. Such an operator resides in the convex space~$\mathcal{S}$ of CPTP operators.

There is another common representation for $\Phi$, which requires the expansion of the quantum-process action $\mathcal{M}[\rho_\textsc{in}]=\sum_{m}K_m\rho_\textsc{in} K^\dag_m$ in terms of some complex Kraus operators $K_m$, where the TP constraint clearly necessitates $\sum_mK^\dag_m K_m=1$. Continuing further, we may in turn expand each $K_m$ with some chosen (non-Hermitian) operator basis $\tr{\Gamma_l^\dag\Gamma_{l'}}=\delta_{l,l'}$, giving us $K_m=\sum^{d^2-1}_{l=0}\gamma_{ml}\Gamma_l$ of complex coefficients $\gamma_{ml}$, and
\begin{equation}
	\rho_\textsc{out}=\sum^{d^2-1}_{l=0}\sum^{d^2-1}_{l'=0}\chi_{ll'}\Gamma_{l}\rho_\textsc{in}\Gamma_{l'}^\dag\,,
	\label{eq:chi_desc}
\end{equation}
where the elements $\chi_{ll'}=\sum_m\gamma_{ml}\gamma_{ml'}^*$. A simplified choice would be $\Gamma_l=\ket{j}\bra{k}$, with $0\leq l=jd+k\leq d^2-1$ for \mbox{$0\leq j,k\leq d-1$}. We may rewrite Eq.~\eqref{eq:chi_desc} as
\begin{align}
	\sket{\rho_\textsc{out}}=&\,\sum^{d^2-1}_{l=0}\sum^{d^2-1}_{l'=0}\chi_{ll'}\sket{\Gamma_{l}\rho_\textsc{in}\Gamma_{l'}^\dag}\nonumber\\
	=&\,\sum^{d^2-1}_{l=0}\sum^{d^2-1}_{l'=0}\chi_{ll'}\Gamma^*_{l'}\otimes\Gamma_l\sket{\rho_\textsc{in}}\equiv\mathcal{\chi}_\Phi^{\top_1}\sket{\rho_\textsc{in}}
	\label{eq:chi_desc2}
\end{align}
using the superket notation~\footnote{An option is the definition $X=\ket{a}\bra{b}\mapsto\sket{X}=\ket{b}\ket{a}$ \emph{via} a \emph{swapped transposition} map on $\bra{b}$. This is equivalent to the vectorization operation $\mathrm{vec}[X]$. Notation-wise, $\sbra{X^\dag}\equiv\sket{X}^\dag$.} and the result $\sket{AXB}=B^\top\otimes A\sket{X}$, and identify the superoperator $\rvec{x}\leftrightarrow\chi_\Phi=\sum^{d^2-1}_{l,l'=0}\chi_{ll'}\Gamma^\dag_{l'}\otimes\Gamma_l$ as a generally non-Hermitian representation of $\Phi$. We find that $\chi_\Phi$ is in fact swap-Hermitian---$\chi_\Phi^\dag=\tau\chi_\Phi\tau$ where $\tau$ is the two-qudit swap operator. A more regular form of this representation can be obtained by noting the equality
\begin{equation}
	\chi_\Phi=\sum_{m}K^\dag_m\otimes K_m
\end{equation}
owing to the superket notation, where the trace-orthonormality of $\Gamma_l$ yields $\chi_{jj'}=\sum_m\tr{\Gamma_{j'} K^\dag_m}\tr{K_m\Gamma^\dag_{j}}$. Since $\tr{A^\dag B}=\sinner{A^\dag}{B}=\Tr{\sket{B}\sbra{A^\dag}}$, 
\begin{equation}
	\chi_{jj'}=\opsinner{\Gamma_j^\dag}{\left(\sum_m\sket{K_m}\sbra{K_m^\dag}\right)}{\Gamma_{j'}}\,,
\end{equation}
so that we may now assign $\widetilde{\chi}_\Phi=\sum^{d^2-1}_{l,l'=0}\sket{\Gamma_{l}}\chi_{ll'}\sbra{\Gamma_{l'}^\dag}=\sum_m\sket{K_m}\sbra{K_m^\dag}$ as a positive superoperator-representation of $\Phi$. For a unitary process $K_1=U$, $\widetilde{\chi}_\Phi=\sket{U}\sbra{U^\dag}$ is another expression of such a rank-one process.

Very often, we are interested in only the elements $\chi_{ll'}$ themselves as they are the unknown parameters of $\Phi$ to be characterized, which may be cleanly packed into a positive operator $\chi=\sum^{d^2-1}_{l,l'=0}\ket{l}\chi_{ll'}\bra{l'}$ with respect to the $d^2$-dimensional computational basis $\{\ket{l'}\}$. The reader can easily verify that $\chi\geq0$. Following~\eqref{eq:chi_desc}, the TP constraint separately states that 
\begin{equation}
	\sum^{d^2-1}_{l,l'=0}\chi_{ll'}\Gamma_{l'}^\dag\Gamma_{l}=1\,,
	\label{eq:cptp_chi}
\end{equation} 
or $\sbra{1}\chi_\Phi^{\top_1}=\sbra{1}$ from \eqref{eq:chi_desc2}~\footnote{Generally speaking, $\chi_\Phi^{\top_1} \sket{1}\neq\sket{1}$ as $\chi_\Phi$ is not a normal operator.}. These properties serve as a reminder that describing $\Phi$ with $\chi$ must always be accompanied by the set of basis operators $\{\Gamma_l\}$, whereas $\chi_\Phi$ already contains all such information implicitly. 

Since $\rho_\Phi$ and $\chi$ describe the same process $\Phi$, these two representations must be unitarily equivalent. To see this, we first define $\ket{e_l}=1\otimes \Gamma_l\ket{\textsc{me}}\sqrt{d}$ and show indeed that these kets form a complete basis:
\begin{align}
	\inner{e_l}{e_{l'}}=&\,d\,\bra{\textsc{me}} 1\otimes \Gamma_l^\dag\Gamma_{l'}\ket{\textsc{me}}\nonumber\\
	=&\,\sum^{d-1}_{j=0}\sum^{d-1}_{j'=0}\inner{j}{j'}\opinner{j}{\Gamma_l^\dag\Gamma_{l'}}{j'}\nonumber\\
	=&\,\tr{\Gamma_{l}^\dag\Gamma_{l'}}=\delta_{l,l'}\,.
\end{align}
It is then a simple matter to realize that indeed,
\begin{equation}
	\rho_\Phi=\sum_{l,l'}\ket{e_{l}}\chi_{ll'}\bra{e_{l'}}=U\chi U^\dag\,,
\end{equation}
where the transformation unitary operator $U=\sum^{d^2-1}_{l=0}\ket{e_l}\bra{l}$ contains information about the basis operators that define $\chi$. Therefore, one might as well interchange between the two descriptions as both representations contain exactly the same information. As a simple example, consider the single-qubit depolarizing channel that is described by the Kraus operators
\begin{align}
	K_1=\sqrt{p_1}\sigma_x\,,&\,\,\,K_2=\sqrt{p_2}\sigma_y,\nonumber\\
	K_3=\sqrt{p_3}\sigma_z\,,&\,\,\,K_4=\sqrt{1-p_1-p_2-p_3}\,.
\end{align}
After some short calculations, one should end up with
\begin{align}
	\rho_\Phi\,\widehat{=}&\begin{pmatrix}
		1-p_1-p_2 & 0 & 0 & 1-p_1-p_2-2p_3\\
		0 & p_1+p_2 & p_1-p_2 & 0\\
		0 & p_1-p_2 & p_1+p_2 & 0\\
		1-p_1-p_2-2p_3 & 0 & 0 & 1-p_1-p_2
	\end{pmatrix}\,,\nonumber\\
	\chi\,\widehat{=}&\begin{pmatrix}
		1-p_1-p_2-p_3 & 0 & 0 & 0\\
		0 & p_1 & 0 & 0\\
		0 & 0 & p_2 & 0\\
		0 & 0 & 0 & p_3
	\end{pmatrix}\,,\nonumber\\
	U\,\widehat{=}&\,\dfrac{1}{\sqrt{2}}\begin{pmatrix}
		\,1 & \hphantom{-}0 & \hphantom{-}0 & \hphantom{-}1\\
		\,0 & \hphantom{-}1 & \hphantom{-}\I & \hphantom{-}0\\
		\,0 & \hphantom{-}1 & -\I & \hphantom{-}0\\
		\,1 & \hphantom{-}0 & \hphantom{-}0 & -1
	\end{pmatrix}\,.
\end{align}

\subsection{Input-state-output-POVM framework}

In this section, we make use of the Choi--Jamio{\l}kowski representation ($\rvec{x}\leftrightarrow\rho_\Phi$) of a given unknown process $\Phi$. Under the usual input-state-output-POVM framework, the relevant true data associated with the measurement, $\rvec{b}\leftrightarrow\{\tr{\rho_\Phi\,\rho_l^\top\otimes\Pi_j}\}$, consist of Born probabilities specified by the set of $L$ input states $\{\rho_l\}^L_{l=1}$ and an $M$-outcome POVM. The linear function $f(\rho')=\tr{\rho'Z}$ for ICC is now defined by a randomly-chosen full-rank state $Z$ of dimension $d^2$, where the convex set $\mathcal{C}$ now specifies the space of CPTP operators $\rho'$. Note that for a CPTP operator, $\ptr{2}{\rho'}=\ptr{2}{\rho_\Phi}=d$.

In this framework, the local structure embedding an input state and a POVM outcome restricts the possibilities of having a direct adaptive method for choosing ``optimal'' measurement configurations to characterize a rank-$r$ operator $\rho_\Phi=\rho_{\Phi,r}$. Nevertheless the POVM (taken to be basis measurements) that probes every output state $\rho_\textsc{out}$ from the unknown $\Phi$ can still be economized using either the ACT or PACT scheme to reduce the number of bases for IC reconstructions. 

In this regard, the (P)ACT-based QPT~[(P)ACTQPT] is a simple iterative scheme that sends a randomly-chosen input state through the unknown process $\rho_{\Phi,r}$ and performs (P)ACT on the output state at every iteration step. After which, ICC \emph{for CPTP operators} is carried out to compute the IC indicator $s_\textsc{cvx}$ from all accumulated output-state physical probabilities at that moment. The scheme repeats itself until $s_\textsc{cvx}=0$. One way to cast the CPTP constraints into a convenient form for the SDPs in ICC is to adopt an earlier methodology~\cite{Audenaert:2002aa} by first employing a set of $d^2-1$ traceless Hermitian basis operators $\{\Omega_{l}\}^{d^2-1}_{l=1}$ that, together with $1/\sqrt{d}$, span both the input and output linear-operator spaces. Under this basis-operator parametrization, it is easy to work out that the $d^2$ equations $\tr{\rho'\,\Omega_{l}\otimes1}=0$ for all $1\leq l\leq d^2-1$ and $\tr{\rho'}=d$ precisely spell the TP constraint $\ptr{2}{\rho'}=d$. Another technical point to note is that the accumulated set of ML probabilities $\{\ML{\rvec{p}}^{(1)},\ML{\rvec{p}}^{(2)},\ldots,\ML{\rvec{p}}^{(L)}\}$ obtained from doing (P)ACT on the output states would generally not be consistent with a CPTP operator. In this case, a second-level regularization can be carried out with the LS procedure that minimizes
\begin{equation}
	g(\{\rvec{p}'_l\})=\sum^{L}_{l=1}\|\rvec{p}'_l-\widehat{\rvec{p}}^{(l)}_{\textsc{ml}}\|^2
\end{equation}
over the CPTP space to acquire a new set of LS probabilities $\{\LS{\rvec{p}}^{(1)},\LS{\rvec{p}}^{(2)},\ldots,\LS{\rvec{p}}^{(L)}\}$ for verifying if the corresponding process reconstruction is unique using ICC. The explicit {\bf Algorithm~\ref{algo:ACTQPT}} that summarizes these steps encapsulates matters presented in a previous article~\cite{Teo:2020cs}.

\begin{figure}[t]
	\centering
	\includegraphics[draft=false,width=1\columnwidth]{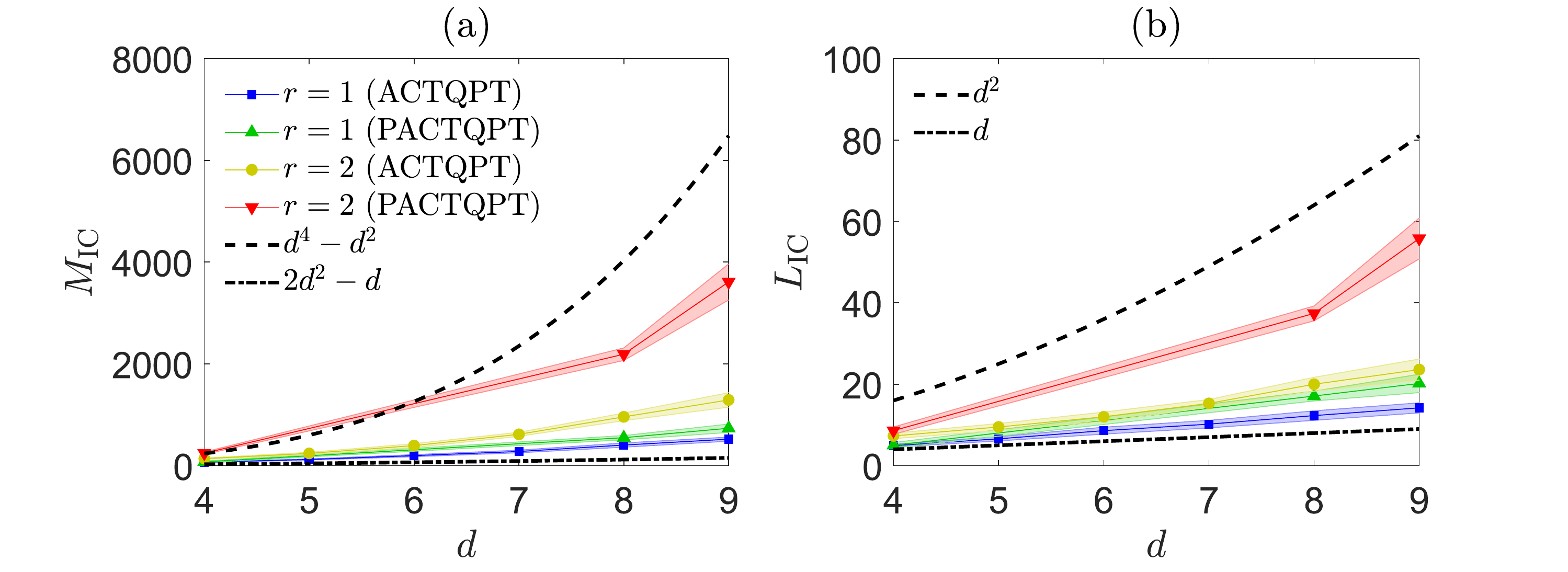}
	\caption{\label{fig:(P)ACTQPT}The (a)~$M_\textsc{ic}$ and (b)~$L_\textsc{ic}$ curves of all noiseless adaptive CQPT schemes on unknown quantum processes of various dimensions. Each 1-$\sigma$ error region represents the statistics of 10 randomly-chosen processes. All PACTQPT schemes are carried out on processes of dimensions $d=4$, 8 and 9. Besides the obvious performance gaps between processes of different ranks, we observe the high compressivity for low-rank processes relative to the respective benchmarks $d^4-d^2$ and $d^2$ for IC QPT obtained from parameter counting. Local input states and product output POVMs that are concurrently employed in PACTQPT yield higher $M_\textsc{ic}$ and $L_\textsc{ic}$ values. When characterizing unitary processes ($r=1$) under the input-state-output-POVM framework, there exists a growing deviation between the entangled ACTQPT and optimal BKD unitary schemes based on projective output-state measurements.}
\end{figure}

Figure~\ref{fig:(P)ACTQPT} highlights the simulated $M_\textsc{ic}$ and $L_\textsc{ic}$ curves for the respective (P)ACTQPT schemes performed on unknown rank-1 and rank-2 quantum processes. These random quantum processes of fixed rank are generated according to~{\bf Algorithm~\ref{algo:RandP}}, with the help of random Kraus operators that are first constructed as building blocks. When the unknown process is unitary ($r=1$), a crucial benchmark for our (P)ACTQPT schemes was developed by Baldwin, Kalev and Deutsch~(BKD)~\cite{Baldwin:2014aa}. For completeness, we reiterate the key points here. Let us first parametrize the unknown unitary operator of interest as $U=\sum^{d-1}_{j=0}|u_j\rangle\langle j|$, where $\{\ket{u_j}\}$ is, in fact, the set of mutually-orthonormal kets we want to characterize, and $\{\ket{j}\}$ is the computational basis. Consider $d$ specific input kets,
\begin{align}
	\ket{\psi_0}=&\,\ket{0}\,,\nonumber\\
	\ket{\psi_l}=&\,(\ket{0}+\ket{l})\dfrac{1}{\sqrt{2}}\,,\quad 1\leq l\leq d-1\,.
	\label{eq:BKDkets}
\end{align}
Thus, sending $\ket{\psi_0}$ to $U$ yields $\ket{u_0}$, and $\ket{\psi_l}$ gives $(\ket{u_0}+\ket{u_l})/\sqrt{2}$. Once $\ket{u_0}$ is completely determined, each subsequent output state requires measurements of fewer than $d$ amplitudes in the computational basis for an IC characterization, the number of which decreases as $l$ increases. For example, the output ket $(\ket{u_0}+\ket{u_l})/\sqrt{2}$ corresponding to the input ket $\ket{\psi_{l}}$, for some $l>0$, can be fully determined by characterizing the amplitudes $\inner{j}{u_{l}}$. However, since all previous $l$ kets in $\{\ket{u_{l'}}\}^{l-1}_{l'=0}$ are determined, we only need to characterize $d-l$ such amplitudes (equivalent to knowing $2d-2l$ real parameters) for $0\leq j\leq d-l-1$ and additionally impose all $2l$ orthogonality relations on all $l$ measured kets $\ket{u_0}$ through $\ket{u_{l-1}}$. The total number of measurements needed is therefore $M_\textsc{ic}=\sum^{d-1}_{l=0}(M_0-2l)=(M_0+1)d-d^2$, where $M_0$ is the number of outcomes required to fully characterize an arbitrary pure state. 

In the original BKD unitary scheme, the authors considered a sophisticated output-state POVM consisting of $M_0=2d$ non-projective~EP~measurement outcomes of the Flammia type~\cite{Flammia:2005fk}, which leads to the main result~$d^2+d$ stated in the article~\cite{Baldwin:2014aa}. We shall, instead, consider only rank-1 projective output-state measurements, which are much more conveniently implementable experimentally. For such measurements, one finds that $M_0\geq3d-2$~\cite{Finkelstein:2004aa}. The corresponding scaling behavior for the \emph{projective} BKD unitary scheme now reads $M_\textsc{ic}=2d^2-d$, with $L_\textsc{ic}=d$ using the specialized input states defined by the kets in~\eqref{eq:BKDkets}. 

In Fig.~\ref{fig:(P)ACTQPT}, we see that for unitary processes, ACTQPT with entangled input states and output POVM still gives $M_\textsc{ic}$ and $L_\textsc{ic}$ that deviate appreciably from the BKD scaling benchmark. Owing to the apparent rigidity of the input-state-output-POVM framework on which (P)ACTQPT schemes are built, an interesting and potentially fruitful question is whether there exists a different CQPT approach that might lead to a better resource-scaling performance beyond such a framework. The answer to this question shall be revealed in the next section.

\subsection{Diagonal-element probing framework}

Solely for the sake of variation, we shall turn to the process-operator $(\chi)$ description of the unknown process $\Phi$ we want to characterize. The motivation for an adaptive CQPT~(ACQPT) scheme, which goes beyond the input-state-output-POVM framework, is triggered by the following trivial observation: for a rank-$r$ CPTP operator $\rvec{x}\leftrightarrow\chi=\chi_r$ of \emph{known} eigenbasis, then only $r-1$ nonzero eigenvalue measurements in its eigenbasis are needed to fully characterize it. The idea of ACQPT is therefore to iteratively acquire the knowledge about $\chi_r$'s eigenbasis with few measurements.

For $d$-dimensional quantum systems, the noiseless measurement $p_M$ at the $M$th iteration step is now the $\kappa_M$th diagonal element of a rotated operator $U_M^\dag\chi_rU_M/d$ with some chosen unitary $U_M$; $p_M=\opinner{\kappa_M}{U_M^\dag\chi_rU_M}{\kappa_M}/d$ for the projector $\ket{\kappa_M}\bra{\kappa_M}$ belonging to the standard computational basis ($0\leq\kappa_M\leq d^2-1$). The choice $U_1=1$ is a sensible neutral choice for $M=1$ as nothing is known about $\chi_r$ at this stage. For $M>1$, $U_{M}$ may be chosen with the help of minENT, which assigns $U_M$ as the eigenbasis of the minENT estimator $\widehat{\chi}_M=\arg\min_{\chi'\in\mathcal{C}_M}S(\chi'/d)$. We bear in mind that the eigenvalues of $\widehat{\chi}_M$ are to be sorted in descending order and accounted for in $U_M$. Given the rank $r_{M-1}$ of $\widehat{\chi}_{M-1}$ from the previous iteration, we may define $\kappa_M=\mod(M-1,r_{M-1})+1$. The motivation for this formula is that as $M$ increases, $\widehat{\chi}_M$ gradually becomes a more accurate reflection of the actual unknown $\chi_r$, so that while the position $\kappa_M$ of the measured diagonal element increases by one with $M$, the rank $r_{M-1}\rightarrow r$ limits the value of $1\leq\kappa_M\leq r_{M-1}$ to within the estimated positive eigenspace of $\chi_r$ where measurements give useful information. 

In the absence of any prior information about $\chi_r$, the aforementioned diagonal-element measurement scheme---ACQPT---can still characterize $\chi$ with a much smaller number of measurements than $O(d^4)$. This scheme is reiterated in {\bf Algorithm~\ref{algo:ACQPT}}. In terms of earlier notations, we have, for the noiseless case, $\rvec{b}=\rvec{p}_\mathrm{diag}$ where $\rvec{p}_\mathrm{diag}=(p_1\,\,p_2\,\,\ldots\,\,p_M)^\top$, and $\dyadic{A}\leftrightarrow \{\ket{\kappa_M}\bra{\kappa_M}\}$. The TP constraint for the ICC routine can be explicitly implemented according to Eq.~\eqref{eq:cptp_chi} with some fixed operator basis $\{\Gamma_l\}^{d^2-1}_{l=0}$.

Furthermore, {\bf Algorithm~\ref{algo:ACQPT}} is universal in the sense that the logic applies for reconstructing any low-rank quantum object $\rvec{x}\leftrightarrow X$ that possesses a convex parameter space $\mathcal{S}$, with the obvious caveat that $\ket{\kappa_M}\bra{\kappa_M}$ must be implementable. In particular, we may also design an ``ACQST'' scheme based on this diagonal-element probing framework, where $\chi=\chi_r$ in {\bf Algorithm~\ref{algo:ACQPT}} is replaced by $\rho=\rho_r$. In this case, we have $p_M=\opinner{\kappa_M}{U_M^\dag\rho_rU_M}{\kappa_M}$, and both $U_M$ and $\ket{\kappa_M}\bra{\kappa_M}$ can be implemented by regular experimental means in a QST experiment. 

On the other hand, a direct estimation of $\opinner{\kappa_M}{U_M^\dag\chi_rU_M}{\kappa_M}$ for a feasible implementation of ACQPT requires more planning. The following line of thought appears to be useful. We first suppose that $U$ diagonalizes $\chi$ into $\Xi$ that houses the eigenvalues, so that the transformation $\Xi=U_M^\dag\chi\,U_M$, or $\chi_{mm'}=\sum_{l}U^{(M)}_{ml}\Xi_{ll}U^{(M)*}_{m'l}$, leads to the transformed action
\begin{equation}
	\mathcal{M}[\rho_\textsc{in}]=\sum^{d^2-1}_{l=0}\Xi_{ll}\,\Gamma'_l\rho_\textsc{in}\Gamma_l'^\dag\,,
	\label{eq:rot_action}
\end{equation}
where $\Xi_{ll}$ are precisely the eigenvalues to be measured and $\Gamma'_l=\sum^{d^2-1}_{m=0}U^{(M)}_{ml}\Gamma_m$ is expressed in terms of the matrix elements of $U_M=\sum_{m,m'}\ket{m}U^{(M)}_{mm'}\bra{m'}$. Measuring the $\kappa_M$th eigenvalue requires specially chosen input state $\rho_\textsc{in}\equiv\rho_M$ and output measurement operator $\Pi_M$ such that $\tr{\Gamma'_l\rho_M\Gamma_l'^\dag\Pi_M}=\delta_{l,\kappa_M}$. If $\Gamma'_l=\ket{j_l}\bra{k_l}$ were all rank-one, where $\inner{j}{k}=\delta_{j,k}$, measuring $\Xi_{\kappa_M\kappa_M}$ can be directly carried out by simply assigning $\rho_M=\ket{k_{\kappa_M}}\bra{k_{\kappa_M}}$ and $\Pi_M=\ket{j_{\kappa_M}}\bra{j_{\kappa_M}}$. Unfortunately, in terms of the singular-value decomposition~(SVD) of $\Gamma_l'=\sum_m\ket{b^{(l)}_m}\lambda^{(l)}_{m}\bra{a^{(l)}_m}$ (ordered in descending $\lambda^{(l)}_{m}$'s), there generally exist several noncommuting rank-one components, which makes assignments of $\rho_M$ and $\Pi_M$ impractical.

Recently~\cite{Kim:2020aa}, we proposed to deploy the largest SV-component approximation $\Gamma'_{\kappa_M}\approx\ket{b_1^{({\kappa_M})}}\bra{a_1^{({\kappa_M})}}\equiv\ket{b_{\kappa_M}}\bra{a_{\kappa_M}}$, where now setting $\rho_M=\ket{a_{\kappa_M}}\bra{a_{\kappa_M}}$ and $\Pi_M=\ket{b_{\kappa_M}}\bra{b_{\kappa_M}}$ approximates the measurement of the $\kappa_M$th eigenvalue. In practice, $U_M$ is clearly never the diagonalizing unitary operator for $\chi$, so the right-hand side of Eq.~\eqref{eq:rot_action} would also include the off-diagonal elements of $U^\dag_M\chi\,U_M\neq\Xi$. Regardless, the largest SV-component approximation can still be used to assign $\rho_M$ and $\Pi_M$ that would eventually become approximately accurate as an eigenvalue-probing asset for larger $M$. The next useful approximation
is to consider product-state approximations of $\rho_M$ and $\Pi_M$ for convenient realization. This may be done by minimizing some distance measure with respect to the $n$-partite unitary operators that rotate $\ket{0}\bra{0}$ to these projectors over the tensor-product spaces of local unitary operators. 

Without repeating the numerical computations, we summarize one relevant result~\cite{Kim:2020aa} reported concerning ACQPT. Figure~1(a) of the Supplemental Material~\footnote{The Supplemental Material is available at \url{https://journals.aps.org/prl/supplemental/10.1103/PhysRevLett.124.210401/ACQPT\_SM.pdf}.} shows important scaling behaviors of various different QPT schemes that include ACQPT for qudit unitary gates (simulated for $2\leq d\leq 7$), where {\bf Algorithm~\ref{algo:ACQPT}} was utilized with the largest SV-component approximation. In particular, we compare ACQPT with the random scheme in which the unknown process was rotated with a Haar-distributed unitary operator. For large $d$, it was extrapolated that the respective scalings for the random scheme and ACQPT are $M_\textsc{ic}\approx O(4.8d^2)$ and $M_\textsc{ic}\approx O(3.5d^2)$. A further improvement was achieved when we impose an additional unitarity assumption. This was done by the simple replacement $r_{M+1}=1$ in step~(8) of {\bf Algorithm~\ref{algo:ACQPT}}. When the unitarity assumption is valid, the resulting ACQPT scaling turns out to be $M_\textsc{ic}\approx O(2.2d^2)$, which is almost identical to the optimal BKD scaling with projective measurements. Note that invoking the rank-one or unitarity assumption does not in any way discount the reliability of ACQPT; if this assumption is wrong, ICC can still terminate the scheme correctly, albeit with more rotated diagonal-element measurements.

\section{Compressive quantum detector tomography}
\label{sec:CQDT}

The third application that is especially important in the field of quantum computation is compressive quantum detector tomography~(CQDT)~\footnote{The ``detector'' symbolizes the POVM.}, where one seeks a resource-efficient scheme to characterize low-rank POVMs (almost always, a set of rank-one projectors in the ideal case) that are always employed in quantum-measurement tasks~\cite{Resch:2007a,Higgins:2007aa,NIELSEN:2003aa,Raussendorf:2001aa,Briegel:2009aa,Zhang:2012bb,Grandi:2017aa,Izumi:2020aa,Maciejewski:2020aa}. We caution the varied notions of QDT, which include the estimation of \emph{certain} physical parameters (such as the quantum efficiency and dark-count rate) describing the given quantum-measurement set~\cite{D'Auria:2011aa}, or the calibration of commuting outcomes~\cite{Chen:2019aa,Lundeen:2009sf,Natarajan:2013bh,Schapeler:2007.16048,Bobrov:2015aa}. In this section, we consider the most general (C)QDT of a set of $M$ positive outcomes $\{\Pi_j\}$ that are mutually noncommutative. Standard QDT without any compressive elements clearly requires $O(d^2)$ mutually linearly-independent input states to characterize the POVM.

The idea of CQDT~\cite{Gianani:2020aa} is similar to other types of quantum compressive schemes. Randomly-chosen input states $\dyadic{A}\leftrightarrow\{\rho_l\}$ are sent to the unknown $M$-outcome POVM ($\rvec{x}\leftrightarrow\{\Pi_j\}$), and the noiseless probabilities $\rvec{b}\leftrightarrow \{p_{jl}=\tr{\rho_l\Pi_j}\}$ are statistically measured through detector counts in terms of the relative frequencies $\sum_{j}\nu_{jl}=1$, which are mapped to physical probabilities $\widehat{b}\leftrightarrow\{\widehat{p}_{jl}\}$ either with ML or LS. It is also easy to see that the POVM constraints---$\Pi_j\geq0$ and $\sum^M_{j=1}\Pi_j=\openone$---amount to a convex POVM space $\mathcal{S}$, to which supplementing the linear physical-probability constraints results in yet another convex set $\mathcal{C}$. 

In an iterative bottom-up scheme, as more input states are fed to the POVM, more information about the latter is acquired from the data and the convex set $\mathcal{C}$ consistent with the physical probability estimators would eventually shrink to a single point---a signature of an IC (C)QDT. This can, once again, be certified using the ICC routine with the physically-mapped probabilities. As the quantum object of interest is now a collection of positive operators, the corresponding linear function for ICC, $f(\{\Pi_j'\})=\sum^M_{j=1}\tr{\Pi'_jZ_j}$, is now specified by a collection of $M$ randomly-chosen full-rank positive operators $Z_j$, which we may conveniently impose to be another fixed POVM. This function is minimized over the space of all $M$-outcome POVMs to compute $s_\textsc{cvx}$. {\bf Algorithm~\ref{algo:CQDT}} provides the complete procedure assuming that LS is employed as the inference method, although any other suitable inference method may be chosen since the POVM estimators should converge to the true POVM for large data samples in the absence of systematic errors.

It turns out that probing an unknown POVM with random pure input states in order to recovery its identity is closely related to the problem of phase retrieval~\cite{ELDAR:2012aa,BANDEIRA:2014aa,Bodmann:2015aa,Xu:2018aa}, where intensity (rank-one) measurements are performed on an unknown Hermitian operator and the resulting expectation values are used to recover the operator. In order to appreciate how closely the two problems are related, we quote the analytical expression of the minimal number of IC input states ($L_\textsc{ic}$) for the latter phase-retrieval problem, which was first conjectured~\cite{ELDAR:2012aa} and later proven~\cite{Xu:2018aa} to be
\begin{equation}
	L_\textsc{ic}=(4dr-4r^2)\eta(\lceil d/2\rceil-r)+d^2\eta(r-\lceil d/2\rceil)\quad\text{(phase retrieval)}\,,
	\label{eq:pr_LIC}
\end{equation}
where $\eta(\,\cdot\,)$ is the usual Heaviside step function. Since a POVM is a set of Hermitian observables and data acquired by the same set of input states can be used to characterize these observables at one go, it is perhaps reasonable to believe that the $L_\textsc{ic}$ for CQDT should also be given by Eq.~\eqref{eq:pr_LIC}. Indeed, this would have been the case if all the outcomes are simply Hermitian. On the contrary, Fig.~2 of our recent CQDT article~\cite{Gianani:2020aa} revealed an interesting twist, that is CQDT always outperforms phase retrieval in terms of the $L_\textsc{ic}$ scaling behavior using arbitrary pure input states for all tested Hilbert-space dimensions and POVM ranks. This numerical finding was given by simulations with randomly-chosen POVMs~(see {\bf Algorithm~\ref{algo:RandPOVM}}) of $M=d^2$, $2d^2$, $3d^2$, with the POVM dimension and rank falling in the respective intervals $2\leq d\leq 10$ and $1\leq r\leq 3$, supplemented with experimental evidence. One can attribute this enhancement to the additional POVM constraints---the requirement that all the Hermitian outcomes must also be positive and sum to the identity---that significantly shrinks the tensor-product linear-operator spaces. Moreover, Fig.~3 of the article shows that even product input states can beat the phase-retrieval scaling, indicating that the influence of the POVM-space boundary strongly induces compression regardless of the kinds of input states. Such high compressivity was witnessed for qubit-power, qutrit-power and other composite dimensions.

\section{Concluding remarks}
\label{sec:conc}

Given an unknown quantum object describing a physical system that we are interested in characterizing, there might be a tendency to impose additional assumptions in attempts to reduce the measurement resources needed to uniquely determine the object. An important conclusion that can be drawn from this review article is that no such assumptions are necessary. The underlying reason is tied to the parameter space of such an object, which is usually bounded by its physical constraints; for quantum states, they are the positivity and unit-trace constraints; for quantum processes, the completely-positive and trace-preserving constraints; all quantum-detector outcomes sum to the identity. The point of modern compressive tomography is the exploitation of the intersection between such a bounded (convex) parameter space and additional measurement-data constraints as a consequence of probing the quantum object. For low-rank objects, the intersection \emph{quickly and verifiably} converges to a single unique point with few measurements.

As explicit examples, we have reviewed three important types of quantum compressive tomography, namely for states, processes and detectors, which are crucial procedures to ensure reliable executions of general quantum tasks. The essential numerical algorithms are listed in \ref{app:algos}. For all these schemes, the reduction in measurement resources appears to be quadratic with respect to the Hilbert-space dimension of the quantum system, which is apparently a basic characteristic originating from the interplay between the measurement and quantum constraints. Although tight analytical bounds on the informationally-complete number of measurements are generally unavailable especially for adaptive schemes, to characterize $d$-dimensional states of rank $r$, numerical evidence suggests that $O(rd)\ll O(d^2)$ POVM outcomes or $O(r)$ von~Neumann measurement bases suffice; $d^2$-dimensional processes of rank $r$ requires at most $O(rd^2)\ll O(d^4)$ measurements; $M$-outcome POVMs of dimension $d$ and rank $r$ need no more than $O(rd)\ll O(d^2)$ input states, where the latter may alternatively be benchmarked with known results from phase-retrieval studies. These order-of-magnitude results are consistent with parameter counting. We conclude by confirming once more all the important properties possessed by these compressive methods that are desirable for their feasible implementation on NISQ devices:

\begin{itemize}
	\item No assumptions about the quantum object are made, besides its dimension~(whether it is a multi-qubit state, a multi-qutrit process or a multi-ququart measurement, \emph{etc.}).
	\item Informational completeness certification of whether some given data are sufficient to uniquely characterize the unknown quantum object can be numerically carried out with polynomial complexities in the dimension with convex optimization. If implementable, such an optimization can be done with exponentially faster computational rates using quantum semidefinite programming~\cite{brando_et_al:LIPIcs:2019:10603}.
	\item Both local and entangled measurements, either random or adaptive, are compressive.
	\item These compressive schemes are immediately implementable through extension from presently-known quantum tomography methods.
\end{itemize}

\section*{Acknowledgments}
\addcontentsline{toc}{section}{Acknowledgments}

The authors are grateful to previous discussions and collaborations with leading experimental groups from the University of Ottawa, Moscow State University, Pohang University of Science and Technology, Roma Tre University and Paderborn University that successfully brought all theoretical results into experimental fruition.

The authors also acknowledge support by the National Research Foundation of Korea (Grant Nos. 2019R1A6A1A10073437, 2019M3E4A1080074, 2020R1A2C1008609, and 2020K2A9A1A06102946), the European Union's Horizon 2020 research and innovation program (ApresSF and STORMYTUNE), and the Ministerio de Ciencia e Innovaci{\'o}n (PGC2018-099183-B-I00).

\appendix

\section{Independent parameters of $\rho_r$}
\label{app:rank-r}

If $\rho_r$ denotes a rank-$r$ quantum state, then it clearly possesses $r$ nonzero eigenvalues by definition. We first note that a pure state $\ket{\,\,\,}\bra{\,\,\,}$ has $2d-2$ independent parameters, as $\ket{\,\,\,}$ may be represented as a column of $2d$ complex numbers with a redundant global phase and one real parameter to be fixed by the unit-trace constraint. With this, the spectral decomposition
\begin{equation}
\rho_r=\sum^{r-1}_{l=0}\ket{\lambda_l}\lambda_l\bra{\lambda_l}
\end{equation}
straightforwardly reveals at most $(2d-2)r+r$ parameters. Taking account that $\tr{\rho_r}=1$, and that the eigenkets $\ket{\lambda_l}$ are mutually orthonormal---$\inner{\lambda_l}{\lambda_{l'}}=\delta_{l,l'}$, we get an additional $r^2-r+1$ linear constraints, giving us a total number of $(2d-2)r+r-(r^2-r)-1=(2d-r)r-1$ independent parameters.

\section{Properties in Sec.~\ref{subsec:ICC}}
\label{app:prop}

We give the reasoning for all the following properties:

\begin{prop}[Noiseless monotonicity]
	\it For noiseless data, suppose that the pairs $\dyadic{A}'$,$\rvec{b}'$ and $\dyadic{A}$,$\rvec{b}$ are such that the former exactly includes the latter. Then $s'_{\textsc{cvx}}\leq s_{\textsc{cvx}}$.  
\end{prop}

Since data are noiseless, we automatically have $\mathcal{C}'\subseteq\mathcal{C}$---any estimator $\rvec{x}_0\in\mathcal{C}'$ satisfying $\dyadic{A}'\rvec{x}_0=\rvec{b}'$ must also satisfy $\dyadic{A}\rvec{x}_0=\rvec{b}$, and is therefore in $\mathcal{C}$. Concavity of $f$ thus implies that $f'_{\mathrm{max}}\leq f_{\mathrm{max}}$ and $f'_{\mathrm{min}}\geq f_{\mathrm{min}}$ so that $s'_{\textsc{cvx}}<s_{\textsc{cvx}}$. Although this property is not guaranteed for noisy data since $\mathcal{C}$ and $\mathcal{C}'$ can be different subspaces, it is still typical for large data samples as the inference mapping $\mathcal{M}$ is statistically consistent~($\widehat{\rvec{b}}$ asymptotically approaches the true data).

\begin{prop}[IC]
	\it For a strictly-concave $f$, $s_\textsc{cvx}=0$ implies that $\mathcal{C}$ contains a single point for \emph{any} $\rvec{b}$.
\end{prop}

The strict concavity of $f$ is crucial here as it imposes unique maximum and minimum points over $\mathcal{C}$. It then follows, regardless of whether $\rvec{b}$ is noisy or not, that if $f_{\mathrm{max}}=f(\widehat{\rvec{x}})=f_{\mathrm{min}}$, $\mathcal{C}=\{\widehat{\rvec{x}}\}$ owing to its convexity. Note that if $f$ is not strictly concave, then it could happen that $f_{\mathrm{max}}=f_{\mathrm{min}}=f(\rvec{x}'_1)=f(\rvec{x}'_2)=\lambda f(\rvec{x}'_1)+(1-\lambda)f(\rvec{x}'_2)$ for any $0\leq\lambda\leq1$ (plateau structure), and so the IC indicator property of $s_{\textsc{cvx}}$ no longer holds.

\begin{prop}[Noisy IC]
	\it Suppose that $\dyadic{A}=(\dyadic{A}_1^\top\,\,\dyadic{A}_2^\top)^\top$ contains two sets of sequentially measured outcomes, first of $\dyadic{A}_1$ and next of $\dyadic{A}_2$, which are accompanied by (noisy) data $\rvec{b}=(\rvec{b}_1^\top\,\,\rvec{b}_2^\top)^\top$. Then any statistically-consistent convex inference map outlined in {\bf Definition~\ref{def:IM}} implies that if $s^{(1)}_{\textsc{cvx}}=0$, it must also be that $s^{(1,2)}_{\textsc{cvx}}=0$ provided that characteristic~(1) is fulfilled for $\dyadic{A}$ and $\rvec{b}$.
\end{prop}

This is a consequence of characteristic~(1) of the function $g$ employed in the convex inference map. Suppose that a tomography experiment is defined by the pair $\dyadic{A},\rvec{b}$. If $s_\textsc{cvx}\neq0$, then by definition, $\rvec{x}_\mathrm{min}=\displaystyle\arg\min_{\rvec{x'}\in\mathcal{S}} g(\dyadic{A}\rvec{x}',\rvec{b})$ is not unique, and $g(\dyadic{A}\rvec{x}',\rvec{b})$ has a plateau structure with the convex domain~$\mathcal{C}$ within which all $\rvec{x}_\mathrm{min}$ yields the same minimum value. Owing to convexity in $g$, this plateau bottom is the only place where $\partial g(\dyadic{A}\rvec{x}',\rvec{b})/\partial\rvec{x}'=\rvec{0}$, and characteristic~(1) of {\bf Definition~\ref{def:IM}} implies that $\mathcal{C}$ contains all $\rvec{x}_\mathrm{min}$'s that satisfy $\dyadic{A}\rvec{x}_\mathrm{min}=\rvec{b}$. 
	
When the accumulated measurements $\dyadic{A}$ become IC, we then have $s_\textsc{cvx}=0$ and this would mean that $g$ no longer has such a plateau structure. As such, since $\rvec{b}$ is noisy, there almost surely exists \emph{no} $\rvec{x}'\in\mathcal{S}$ for which $\dyadic{A}\rvec{x}'=\rvec{b}$, so that $\displaystyle\rvec{x}_\mathrm{min}=\arg\min_{\rvec{x'}\in\mathcal{S}} g(\dyadic{A}\rvec{x}',\rvec{b})$ is unique and typically non-stationary.
	
Now suppose the sequential measurements $\dyadic{A}=(\dyadic{A}_1^\top\,\,\dyadic{A}_2^\top)^\top$ were performed, where $\rvec{b}=(\rvec{b}_1^\top\,\,\rvec{b}_2^\top)^\top$. Upon denoting the respective minimum points of the convex inference map for $\dyadic{A}_1,\rvec{b}_1$ and $\dyadic{A},\rvec{b}$ by $\rvec{x}^{(1)}_\mathrm{min}$ and $\rvec{x}^{(1,2)}_\mathrm{min}$, the assertion $s^{(1)}_\textsc{cvx}=0$ means that $\rvec{x}^{(1)}_\mathrm{min}\in\mathcal{S}$ is unique. For such an IC scenario, we first consider the hypothetical noiseless case where $\rvec{x}^{(1)}_\mathrm{min}=\rvec{x}$, the true quantum object. One can either simply invoke {\bf Property~\ref{prop:monotone}}, or trivially see that $s^{(1,2)}_\textsc{cvx}=0$, otherwise $\rvec{x}^{(1)}_\mathrm{min}$ cannot be unique. For \emph{noisy IC scenarios}, we usually have $\dyadic{A}_1\rvec{x}^{(1)}_\mathrm{min}\neq\rvec{b}_1$. A problem shall also arise if $s^{(1,2)}_\textsc{cvx}\neq0$, for this implies that $\dyadic{A}\rvec{x}^{(1,2)}_\mathrm{min}=\rvec{b}$, and so $\rvec{x}^{(1,2)}_\mathrm{min}$ should have been the previous minimum. Even if $\dyadic{A}_1\rvec{x}^{(1)}_\mathrm{min}=\rvec{b}_1$,  $s^{(1,2)}_\textsc{cvx}\neq0$ would again contradict the uniqueness property of $\rvec{x}^{(1)}_\mathrm{min}$.

\section{Compression induced by the state-space boundary}
\label{app:bd_thm}

Suppose that the eigenbasis $\{\ket{\lambda_j}\}$ of a rank-$r$ 
\begin{equation}
	\rho=\rho_r=\sum^{r-1}_{j=0}\ket{\lambda_j}p_j\bra{\lambda_j}
\end{equation}
is measured and the probabilities $\sum^{r-1}_{j=0}p_j=1$ are collected. At this point, the estimator $\widehat{\rho}=\rho_r$ is \emph{one} solution that is consistent with the Born probabilities $\opinner{\lambda_j}{\widehat{\rho}}{\lambda_j}=p_j$. Hermiticity then imposes the general form 
\begin{equation}
	\widehat{\rho}=\rho_r+\sum^{r-1}_{j\neq k=0}\ket{\lambda_j}c_{jk}\bra{\lambda_k}+W
	\label{eq:rhoest}
\end{equation}
for $\widehat{\rho}$, where $W$ is a traceless Hermitian operator outside the support of $\rho$ ($W\rho=0=\rho\,W$). Moreover, $p_{j\geq r}=0=\opinner{\lambda_{j\geq r}}{\widehat{\rho}}{\lambda_{j\geq r}}$ implies that $W$ is represented by a hollow (all diagonal entries equal to zero) Hermitian matrix in the basis $\{\ket{\lambda_{j\geq r}}\}$ with arbitrary off-diagonal entries; the corresponding Hermitian solution subspace for $\widehat{\rho}$ has a nonzero volume. 

It is now obvious that if $\ket{\phi}$ is the eigenket of $W$ that gives a negative eigenvalue, one expects $\opinner{\phi}{\widehat{\rho}}{\phi}=\opinner{\phi}{W}{\phi}<0$. In other words, any such nonzero traceless $W$ always leads to a nonpositive $\widehat{\rho}$. If the quantum positivity constraint is imposed on the solution for the Born probabilities, we must necessarily have $W=0$. 

This leaves the operator $A=\sum^{r-1}_{j\neq k=0}\ket{\lambda_j}c_{jk}\bra{\lambda_k}$ in the right-hand side of Eq.~\eqref{eq:rhoest}. For pure states $(r=1)$, $A$ is clearly zero, so that $\widehat{\rho}=\rho_1=\ket{\lambda_1}\bra{\lambda_1}$ is the unique positive estimator after measuring $\rho_1$ as we expect. For $1<r\leq d$, we note that if $\widehat{\rho}$ is to be consistent with probabilities $\opinner{w_j}{\widehat{\rho}}{w_j}=\opinner{w_j}{\rho_r}{w_j}$ obtained from any other orthonormal measurement basis $\{\ket{w_j}\}$, then $\opinner{w_j}{A}{w_j}=0$ for all $1\leq j\leq d-1$. As $A$ has exactly $r^2-r$ free parameters, it follows that measuring $K_0=\lceil(r^2-r)/(d-1)\rceil+1$ linearly independent bases results in the unique solution $A=0$ to $K_0(d-1)$ \mbox{(pseudo-)}invertible linear equations. Evidently, $K_0=2$ if $r\ll d$.

\newpage
\section{Algorithms}
\label{app:algos}

This Appendix section consolidates all algorithms in order of their respective initial discussions.

\begin{minipage}[c][6cm][c]{0.925\columnwidth}
	\noindent
	\begin{flushleft}
		{\bf Algorithm~\algoct: ICC}\algolab{algo:ICC}\vspace{-2ex}
	\end{flushleft}
	\rule{\columnwidth}{1.5pt}\\
	Define $f(\rvec{x}')=\rvec{z}^\top\rvec{x}'$ for a fixed, randomly-chosen $\rvec{z}$ such that $f$ has no plateau structure.
	\begin{enumerate}
		\item Use a convex inference method (or otherwise) to map the noisy data $\rvec{b}$ to $\widehat{b}$.
		\item Maximize and minimize $f(\rvec{x}')$ to obtain $f_{\mathrm{max}}$ and $f_{\mathrm{min}}$
		subject to
		\begin{itemize}
			\item $\rvec{x}'\in\mathcal{S}$\,,
			\item $\dyadic{A}\rvec{x}'=\widehat{\rvec{b}}'$\,.
		\end{itemize}
		\item Compute $s_{\textsc{cvx}}=f_\mathrm{max}-f_\mathrm{min}$.\\[-5ex]
	\end{enumerate}
	\rule{\columnwidth}{1.5pt}
\end{minipage}

\begin{minipage}[c][6.5cm][c]{0.925\columnwidth}
	\noindent
	\begin{flushleft}
		{\bf Algorithm~\algoct: Random-bases CQST}\algolab{algo:HaarCQST}\vspace{-2ex}
	\end{flushleft}
	\rule{\columnwidth}{1.5pt}\\
	Pre-choose $\varepsilon$ as some numerically small number. Then, beginning with $K=1$ and the standard computational basis $\mathcal{B}_1$:
	\begin{enumerate}
		\item Collect data for $\mathcal{B}_K$ and define the accumulated relative-frequency data $\rvec{\nu}=(\rvec{\nu}_1^\top\,\,\rvec{\nu}_2^\top\,\,\ldots\,\,\rvec{\nu}_K^\top)^\top$.
		\item Perform ML using $\rvec{\nu}$ to obtain $\ML{\rvec{p}}$.
		\item Carry out ICC with $\ML{\rvec{p}}$ to compute $s_{\textsc{cvx},K}$.
		\item If $s_{\textsc{cvx},K}<\varepsilon$ ($K=K_\textsc{ic}$), terminate this scheme and obtain the unique estimator $\widehat{\rho}$ from ICC. Otherwise, select the next basis $\mathcal{B}_{K+1}$, which can either be entangled or local.
		\item Raise $K$ by 1 and repeat all the above steps.\\[-5ex]
	\end{enumerate}
	\rule{\columnwidth}{1.5pt}
\end{minipage}

\begin{center}
	\begin{minipage}[c][5cm][c]{0.925\columnwidth}
		\noindent
		\begin{flushleft}
			{\bf Algorithm~\algoct: Haar-distributed basis construction}\algolab{algo:HaarB}\vspace{-2ex}
		\end{flushleft}
		\rule{\columnwidth}{1.5pt}\\
		Let $\mathcal{B}_\text{ref}=\{\ket{0},\ket{1},\ldots,\ket{d-1}\}$:
		\begin{enumerate}
			\item Generate a random $d\times d$ matrix $\dyadic{A}$ with entries i.i.d. standard Gaussian distribution.
			\item Compute $\dyadic{Q}$ and $\dyadic{R}$ from the QR decomposition $\dyadic{A}=\dyadic{Q}\dyadic{R}$.
			\item Define $\dyadic{R}_\text{diag}=\mathrm{diag}\{\dyadic{R}\}$, $\dyadic{L}=\dyadic{R}_\text{diag} \oslash |\dyadic{R}_\text{diag}|$ and $\dyadic{U}_\mathrm{Haar}=\dyadic{Q}\dyadic{L}$, where~$\oslash$ denotes the Hadamard (entrywise) division.
			\item Construct the Haar-distributed basis $\mathcal{B}_\mathrm{Haar}=\dyadic{U}_\mathrm{Haar}\mathcal{B}_\text{ref}$.\\[-5ex]
		\end{enumerate}
		\rule{\columnwidth}{1.5pt}
	\end{minipage}
\end{center}

\begin{center}
	\begin{minipage}[c][4cm][c]{0.925\columnwidth}
		\noindent
		\begin{flushleft}
			{\bf Algorithm~\algoct: Random rank-$r$ quantum state construction}\algolab{algo:randstate}\vspace{-2ex}
		\end{flushleft}
		\rule{\columnwidth}{1.5pt}
		\begin{enumerate}
			\item Generate a random $r\times d$ matrix $\dyadic{M}$ with entries i.i.d. standard Gaussian distribution.
			\item Define the quantum state $\rho\,\widehat{=}\,\dfrac{\dyadic{M}^\dag \dyadic{M}}{\tr{\dyadic{M}^\dag \dyadic{M}}}$.\\[-4ex]
		\end{enumerate}
		\rule{\columnwidth}{1.5pt}
	\end{minipage}
\end{center}

\begin{minipage}[c][8cm][c]{0.925\columnwidth}
	\noindent
	\begin{flushleft}
		{\bf Algorithm~\algoct: (P)ACT (minENT)}\algolab{algo:ACT}\vspace{-2ex}
	\end{flushleft}
	\rule{\columnwidth}{1.5pt}\\
	Let $\varepsilon$ be some numerically small number. Then, beginning with $K=1$ and the standard computational basis $\mathcal{B}_1$:
	\begin{enumerate}
		\item Collect data $\rvec{\nu}_K$ for $\mathcal{B}_K$ and define the accumulated relative-frequency data $\rvec{\nu}=(\rvec{\nu}_1^\top\,\,\rvec{\nu}_2^\top\,\,\ldots\,\,\rvec{\nu}_K^\top)^\top$.
		\item Perform ML using $\rvec{\nu}$ to obtain $\ML{\rvec{p}}$.
		\item Carry out ICC with $\ML{\rvec{p}}$ to compute $s_{\textsc{cvx},K}$.
		\item If $s_{\textsc{cvx},K}<\varepsilon$ ($K=K_\textsc{ic}$), terminate this scheme and obtain the unique estimator $\widehat{\rho}$ from ICC. Otherwise, proceed.
		\item Minimize $S(\rho')=-\tr{\rho'\log\rho'}$ over $\mathcal{C}_K$, and let $\mathcal{B}_\text{minENT}$ be the eigenbasis of the resulting minENT estimator.
		\begin{itemize}
			\item For ACT, pick $\mathcal{B}_{K+1}=\mathcal{B}_\text{minENT}$.
			\item For PACT, pick $\mathcal{B}_{K+1}$ as the nearest local basis from $\mathcal{B}_\text{minENT}$.
		\end{itemize}
		\item Raise $K$ by 1 and repeat all the above steps.\\[-5ex]
	\end{enumerate}
	\rule{\columnwidth}{1.5pt}
\end{minipage}

\begin{minipage}[c][7cm][c]{0.925\columnwidth}
	\noindent
	\begin{flushleft}
		{\bf Algorithm~\algoct: (P)ACTQPT}\algolab{algo:ACTQPT}\vspace{-2ex}
	\end{flushleft}
	\rule{\columnwidth}{1.5pt}\\
	Let $\varepsilon$ be some numerically small number. Then, beginning with $L=1$:
	\begin{enumerate}
		\item Send a randomly-generated input state $\rho_L$ through the unknown process~$\Phi$.
		\item Perform (P)ACT on the output state and obtain $\ML{\rvec{p}}^{(L)}$. Append this to the already existing physical datasets, $\mathbb{D}_L=\{\ML{\rvec{p}}^{(1)},\ML{\rvec{p}}^{(2)},\ldots,\ML{\rvec{p}}^{(L)}\}$. 
		\item Carry out a second-level regularization by performing a LS procedure on $\mathbb{D}_L$ to obtain $\mathbb{D}'_L=\{\LS{\rvec{p}}^{(1)},\LS{\rvec{p}}^{(2)},\ldots,\LS{\rvec{p}}^{(L)}\}$ over the CPTP space. 
		\item Carry out ICC over the CPTP space using $\mathbb{D}'_L$ to compute $s_{\textsc{cvx},L}$.
		\item If $s_{\textsc{cvx},L}<\varepsilon$ ($L_\textsc{ic}=L$), terminate this scheme and obtain the unique $\widehat{\rho}_\Phi$ from ICC. Otherwise, raise $L$ by 1 and repeat all the above steps.\\[-5ex]
	\end{enumerate}
	\rule{\columnwidth}{1.5pt}
\end{minipage}

\begin{minipage}[c][6cm][c]{0.925\columnwidth}
	\noindent
	\begin{flushleft}
		{\bf Algorithm~\algoct: Random rank-$r$ quantum process construction}\algolab{algo:RandP}\vspace{-2ex}
	\end{flushleft}
	\rule{\columnwidth}{1.5pt}
	\begin{enumerate}
		\item Generate $r$ random $d\times d$ complex matrices $\{\dyadic{M}_m\}^r_{l=1}$, where $\dyadic{M}_m$ has entries i.i.d. standard Gaussian distribution. These arbitrary matrices are linearly-independent so long as $r\leq d$.
		\item Set $\dyadic{S}=\sum_{m=1}^r\dyadic{M}_m^\dag\dyadic{M}_m$.
		\item Define the Kraus operators $K_m\,\widehat{=}\,\dyadic{M}_m \dyadic{S}^{-1/2}$.
		\item With these Kraus operators, one may simulate experiments of a rank-$r$ process using either the Choi-Jamio{\l}kowski or process-operator formalism.\\[-5ex]
	\end{enumerate}
	\rule{\columnwidth}{1.5pt}
\end{minipage}

\begin{minipage}[c][9.5cm][c]{0.925\columnwidth}
	\noindent
	\begin{flushleft}
		{\bf Algorithm~\algoct: ACQPT (ideal)}\algolab{algo:ACQPT}\vspace{-2ex}
	\end{flushleft}
	\rule{\columnwidth}{1.5pt}\\
	Let $\varepsilon$ be some numerically small number. Then, beginning with $M=1$, $U_1=1$ and $\kappa_1=1$:
	\begin{enumerate}
		\item Rotate the unknown $\chi$ to $U_M^\dag\chi_rU_M$.
		\item Measure the two-outcome POVM $\{\ket{\kappa_M}\bra{\kappa_M},1-\ket{\kappa_M}\bra{\kappa_M}\}$ and collect the corresponding normalized relative frequencies $\{\nu_M,1-\nu_M\}$. Define the accumulated data $\rvec{\nu}=(\nu_1\,\,\nu_2\,\,\ldots\,\,\nu_M)^\top$. 
		\item Perform ML to map $\rvec{\nu}$ to $\ML{\rvec{p}}$.
		\item Carry out ICC over the CPTP space of $\chi$ using $\ML{\rvec{p}}$ to compute $s_{\textsc{cvx},M}$.
		\item If $s_{\textsc{cvx},M}<\varepsilon$ ($M_\textsc{ic}=M$), terminate this scheme and obtain the unique $\widehat{\chi}$ from ICC. Otherwise, proceed.
		\item Minimize $S(\chi'/d)=-\tr{(\chi'/d)\log(\chi'/d)}$ over $\mathcal{C}_M$, and obtain the minENT estimator $\widehat{\chi}_\text{minENT}$.
		\item Find the diagonalizing unitary operator $U_{M+1}$ that corresponds to an \emph{ordered} eigenspectrum in descending order. Define $r_{M+1}$ as the rank of 
		\item Set $\kappa_{M+1}=\mod(M,r_{M+1})+1$.
		\item Raise $M$ by 1 and repeat all the above steps.\\[-5ex]
	\end{enumerate}
	\rule{\columnwidth}{1.5pt}
\end{minipage}

\begin{minipage}[c][6.5cm][c]{0.925\columnwidth}
	\noindent
	\begin{flushleft}
		{\bf Algorithm~\algoct: CQDT}\algolab{algo:CQDT}\vspace{-2ex}
	\end{flushleft}
	\rule{\columnwidth}{1.5pt}\\
	Pre-choose $\varepsilon$ as some numerically small number. Then, beginning with $L=1$:
	\begin{enumerate}
		\item Randomly choose an input state $\rho_L$.
		\item Collect data for this input state and define the accumulated relative-frequency data $\rvec{\nu}=(\rvec{\nu}_1^\top\,\,\rvec{\nu}_2^\top\,\,\ldots\,\,\rvec{\nu}_L^\top)^\top$.
		\item Perform LS using $\rvec{\nu}$ to obtain $\LS{\rvec{p}}$.
		\item Carry out ICC with $\LS{\rvec{p}}$ to compute $s_{\textsc{cvx},L}$.
		\item If $s_{\textsc{cvx},L}<\varepsilon$ ($L=L_\textsc{ic}$), terminate this scheme and obtain the unique POVM estimator~$\{\widehat{\Pi_j}\}$ from ICC. Otherwise, select the next input state $\rho_{L+1}$.
		\item Raise $L$ by 1 and repeat all the above steps.\\[-5ex]
	\end{enumerate}
	\rule{\columnwidth}{1.5pt}
\end{minipage}

\begin{minipage}[c][5cm][c]{0.925\columnwidth}
	\noindent
	\begin{flushleft}
		{\bf Algorithm~\algoct: Random rank-$r$ POVM construction}\algolab{algo:RandPOVM}\vspace{-2ex}
	\end{flushleft}
	\rule{\columnwidth}{1.5pt}
	\begin{enumerate}
		\item Generate $M$ random $r\times d$ complex matrices $\{\dyadic{M}_m\}^M_{l=1}$, where $\dyadic{M}_m$ has entries i.i.d. standard Gaussian distribution.
		\item Set $\dyadic{S}=\sum_{m=1}^r\dyadic{M}_m^\dag\dyadic{M}_m$.
		\item Define the POVM outcomes as $\Pi_m\,\widehat{=}\,\dyadic{S}^{-1/2}\dyadic{M}_m^\dag\dyadic{M}_m \dyadic{S}^{-1/2}$.\\[-5ex]
	\end{enumerate}
	\rule{\columnwidth}{1.5pt}
\end{minipage}

\renewcommand\bibname{References}

\bibliographystyle{ws-ijqi}

\end{document}